\documentclass[a4paper,11pt]{article}
\pdfoutput=1 

\usepackage{jheppub} 

\usepackage[T1]{fontenc} 

\usepackage{scalerel}
\usepackage{color}
\usepackage{latexsym}
\usepackage{mathrsfs}
\usepackage{amssymb}
\usepackage{amsmath}
\usepackage{amsfonts, epsf,epsfig,color}
\usepackage{graphicx}
\usepackage{tensor}
\usepackage{manfnt}
\usepackage{slashed}
\usepackage[all]{xy}
\usepackage[]{tensor}
\include{macro}

\newcommand{\R}{\mathbb{R}}

\newcommand{\Z}{\mathbb{Z}}

\newcommand{\rd}{\, \mathrm{d}}

\newcommand{\be}{\begin{equation}\label}
\newcommand{\ee}{\end{equation}}
\newcommand{\bea}{\begin{eqnarray}\label}
\newcommand{\eea}{\end{eqnarray}}

\title{\textbf{Torus Bundles, Automorphisms and T-Duality}}

\author{H. Mahmood and R. A. Reid-Edwards}

\affiliation{Department of Applied Mathematics and Theoretical Physics \\ University of Cambridge, Cambridge, CB3 0WA, UK}

\emailAdd{hm516@cam.ac.uk}
\emailAdd{r.a.reid-edwards@damtp.cam.ac.uk}

\abstract{We reconsider some older constructions of T-duality, based on automorphisms of the worldsheet operator algebra, in a modern context. It has been long known that at special points in the moduli space of torus compactifications, the target space gauge symmetry may be enhanced. Away from such points the symmetry is broken and T-duality may be understood as a residual discrete gauge symmetry that survives this breaking. Drawing on work on connections over the space of string backgrounds, we discuss how to generalise this framework for T-duality to geometric and non-geometric backgrounds that are not full solutions of string theory, but may play an important role in exact backgrounds. Along the way we find an interesting algebraic structure and discuss its relationship with doubled geometry. We comment on non-isometric T-duality in this context.
}

\begin{document} 
\maketitle

\flushbottom

\section{Introduction}

The most useful way of proving that two apparently different theories are really just different T-dual descriptions of the same theory is the isometry gauging procedure originally introduced by Buscher in \cite{Buscher:1987sk} and further refined in \cite{Buscher:1987qj,Hull:2006qs}. The validity of the Buscher procedure rests on the existence of a compact abelian isometry of the target space which can be used to generate a rigid symmetry of the worldsheet theory. This symmetry can then be gauged in the sigma model and the gauge fields then integrated out to give a dual description of the gauged theory. If one can then show that the gauged theory is equivalent to the ungauged theory, then one has a pair of worldsheet theories that describe the same physics.  There are many cases where the required symmetry of the target space does not exist (or may exist only locally), but there is still evidence that a dual description of the theory exists and there is a sense that the relationship between these two descriptions should still be thought of as a form of T-duality. Examples include the SYZ description of Mirror Symmetry \cite{Strominger:1996it} that involves torus bundles where the fibres can degenerate, as well as T-folds \cite{Hull:2004in}, whose non-geometric nature can be traced to an obstruction to extending a local circle isometry globally in the geometric dual \cite{Hull:2009sg}.

Progress has been made in casting the Buscher construction in more general (and exotic) settings \cite{Aldazabal:2013sca}, but we always fall back to the same restrictions when attempting to prove the duality rigorously, at the level of the worldsheet theory. This stands in contrast to the remarkable progress that has taken place in incorporating duality symmetries directly into supergravity-inspired field theory constructions\footnote{See \cite{Berman:2013eva} for a nice review of some of the more recent advances.}. In this paper, instead of attempting to generalise the Buscher construction, we revisit the perspective of T-duality as an automorphism of the operator algebra. Using ideas discussed in \cite{Evans:1995su} and with a view of moving beyond the isometric torus bundle paradigm, whilst remaining within the context of a symmetry that is recognisable as T-duality, we reconsider this operator algebra approach in a contemporary context. We will not settle the issue of non-isometric T-duality in this paper, but we hope that the constructions presented and the observations made will help to indicate a possible alternative approach, one which we feel has not yet been fully explored.

\subsection{Symmetries and Automorphisms}

The starting point is the key observation of \cite{Dine:1989vu} that T-duality on a circle may be understood as a residual discrete symmetry that endures after the breaking of a larger enhanced gauge symmetry. The larger symmetry is only manifest for special backgrounds and it is in this sense that T-duality should be thought of as a gauge symmetry of the target space. Such gauge symmetries manifest themselves as automorphisms of the operator algebra of the worldsheet theory. Given a conserved charge $Q$, we can use it to act with the operator algebra of a string theory as an (inner) automorphism
\begin{equation}\label{aut}
{\cal A}\rightarrow e^{iQ}{\cal A} e^{-iQ},
\end{equation}
where ${\cal A}$ is an operator in the worldsheet theory. Now, any transformation of this kind will be a symmetry, but those transformations that map a recognisable theory to an apparently different theory of the same kind will be of particular interest. For example, if the operator ${\cal A}$ is the stress tensor of a string theory and $Q$ is such that the automorphism maps it to a stress tensor of an apparently different string theory, then the symmetry can be thought of as a string-string duality. The particular case where $Q$ generates a T-duality will be the subject of this paper.

The automorphism (\ref{aut}) is a symmetry of the theory and so any infinitesimal deformation of the form
\begin{equation}
\delta {\cal A}=i[Q,{\cal A}]
\end{equation}
gives a symmetry of the theory. This can be contrasted with general transformations, generated by vertex operators, that correspond to genuine physical deformations of the theory. This description of gauge symmetry, explored in \cite{Evans:1989xq}, is elegantly encoded in the BRST framework of String Field Theory, wherein symmetries of the target space are generated by BRST transformations of the string field\footnote{See \cite{Zwiebach:1992ie} for details.}
\begin{equation}
\delta|\Psi\rangle=Q_B|\Lambda\rangle+...,
\end{equation}
where $+...$ denotes non-linear terms. Obvious candidates for such conserved charges are those of the form
\begin{equation}
Q=\oint\frac{\rd z}{2\pi i}\,\Lambda\, J(z),
\end{equation}
where $J(z)$ is a weight $(1,0)$ holomorphic worldsheet current, or the obvious anti-holomorphic counterpart, or combinations of both. Examples include $B$-field transformations and diffeomorphisms of the target space, where the appropriate conserved currents are $J=\xi_i\partial X^{i }(z)+\zeta_i\bar{\partial}X^{i}(\bar{z})$. Questions of conservation (commutation with the worldsheet Hamiltonian) and the fact that the natural symmetries from the target space perspective can be simply understood in terms of the combinations $X^{i }_R(\bar{z})\pm X^{i }_L(z)$, rather than $X^{i }_R(\bar{z})$ and $X^{i }_L(z)$ separately, indicate that the most natural language in which to discuss these symmetries is the canonical one in which the worldsheet has a Lorentzian metric.

As discussed above, at special points in the moduli space of string backgrounds we see an enhancement of the target space symmetry as additional states become massless and form multiplets of non-abelian gauge symmetries. The classic example of this in the closed string sector is the Halpern-Frenkel-Kac-Segal (HFKS) mechanism \cite{Frenkel:1980rn,Segal:1981ap,Halpern:1975nm} in which the $d$ commuting currents $H^i=i\partial X_L^i(z)$, where $i=1,..d$, are joined by the currents
$E_{\pm\alpha}=:e^{\pm i\alpha_iX^i_L(z)}:$, which are weight $(1,0)$ at this enhancement point in the moduli space. The $\alpha_i$ are the root vectors  of the enhanced group, which has rank $d$. One perspective on this \cite{Kugo:1992md} is that the enhanced symmetry is a gauge symmetry of the background independent theory that is generically broken by a choice of background.

\subsection{T-Duality as a Gauge Symmetry}

An example of such HFKS enhancement is when the target space includes a circle at the self-dual radius $R=\sqrt{\alpha'}$, i.e. $d=1$, and the $U(1)_L\times U(1)_R$ symmetry generated by the currents $\partial X\pm\bar{\partial}X$ is enhanced to $SU(2)_L\times SU(2)_R$ - one copy of $SU(2)$ each for the chiral and anti-chiral sectors. The currents $\partial X$, $e^{\pm i2X_L/\sqrt{\alpha'}}$ generate $SU(2)_L$, with similar right-moving currents generating the $SU(2)_R$. It was shown in \cite{Dine:1989vu} that away from the self-dual radius, the gauge symmetry breaks to the Cartan $U(1)\times U(1)$ with a residual discrete $\Z_2$ gauge symmetry which could be identified with T-Duality in the circle. \cite{Evans:1995su} showed that the charge responsible for the action of the $\Z_2\subset SU(2)_L$,
\begin{equation}
Q=\frac{1}{2}\oint\rd\sigma \sin\left(2X_L(\sigma)\right),
\end{equation}
could still be used to generate the T-duality transformation away from the self-dual radius, even though the current $\sin(2X_L(\sigma))$ is generally not conserved\footnote{We use conventions where the coordinates are dimensionless, so the charge looks different to that given in \cite{Evans:1995su}. Note that all factors of $\alpha'$ are absorbed into the metric and B-field, so the coordinates are dimensionless.}. The key was to write the fields of the theory at radius $R$ in terms of the fields defined at the self-dual radius. This then ensured that the action of the charge $Q$ could be computed on fields  defined away from the self-dual radius. Thus the effect of an automorphism, by this charge, on the operator algebra of the theory at generic radius could be computed. It was shown in \cite{Evans:1995su} that this procedure correctly reproduces the Buscher rules. The extension of this framework to the larger $O(d,d;\Z)$ symmetry group was discussed in \cite{Giveon:1990era} and demonstrated explicitly for $d=2$ in \cite{Giannakis:1996sv}. Importantly, if 
\begin{equation}
Q_{\Lambda} = \Lambda\oint \rd\sigma \sin(2X_L(\sigma)),
\end{equation}
the current is not of weight $(1,0)$ and this charge will not be conserved for general values of $\Lambda$. Surprisingly, the automorphism still makes sense away from the self-dual radius if $\Lambda=\frac{1}{2}$. We shall discuss why this is the case in section \ref{s: an algebraic approach to T-duality}.

The main result of \cite{Evans:1995su} stems from the fact that, at the self-dual radius,
\begin{equation}\label{transformations}
e^{iQ}\partial X(\sigma)e^{-iQ}=-\partial X(\sigma),    \qquad e^{iQ}\bar{\partial} X(\sigma)e^{-iQ}=\bar{\partial} X(\sigma).
\end{equation}
Away from the self-dual radius, the transformation is more complicated. If one knows how to write the fields of the theory at a particular background, e.g. at $R=\sqrt{\alpha'}$, in terms of the Hilbert space of another background, then (\ref{transformations}) can be used to deduce the duality transformations. We can define a basis for the operators at a given point. Most of our considerations will involve the operators constructed from combinations of $\partial X$ and $\bar{\partial}X$. The natural way to do this is to define a connection on the space of backgrounds and then to parallel transport, with respect to that connection, the basis of states or operators at a point of enhanced symmetry to the background of interest. This may sound like a tall order, but some progress has been made on this general issue \cite{Campbell:1990dz,Kugo:1992md,Ranganathan:1993vj,Ranganathan:1992nb,Sen:1993mh} and, as we shall see later, this issue simplifies greatly in a certain class of backgrounds.

\subsection{Outline}

The outline of this paper is as follows: In section \ref{s: an algebraic approach to T-duality} we review the algebraic approach to T-duality pioneered in \cite{Dine:1989vu} and then further developed in \cite{Evans:1995su}. We discuss the role of universal coordinates \cite{Kugo:1992md}, objects that may have a natural meaning at all points on the space of torus bundle backgrounds, and show that they are the natural framework for an algebraic notion of T-duality for general torus bundles as an (inner) automorphism of the operator algebra. We take the opportunity to further develop and to clarify various issues that may not have been fully appreciated previously or that we could not find discussion of in the literature. We also take this as an opportunity to highlight work in the older literature that we feel may be relevant to contemporary discussions, but may have been overlooked.

In section \ref{s: torus bundles} we outline a more general framework, in which a connection is used to transport the Hilbert space from a point of symmetry enhancement to a particular background of interest. Using work pioneered in \cite{Ranganathan:1992nb,Ranganathan:1993vj} on the connection on the space of CFTs, we outline a framework for generalising the operator algebra automorphism of T-duality to more general backgrounds. We feel this framework has the potential to cover a much larger class of torus bundle backgrounds than the conventional Buscher procedure, possibly including cases where fibres degenerate; however, our limited knowledge of the space of string backgrounds makes this framework practical in only select cases. We stress that the framework is applicable to sigma models, not just CFTs, thus making the framework applicable to toy models, not just string backgrounds with a bona fide CFT.

In section \ref{s: torus bundle examples} we apply this framework in the familiar context of torus bundles over circles with monodromy and show how the conventional nilfold, torus with $H$-flux and T-fold sigma models \cite{Hull:2009sg, DallAgata:2007egc} appear in this framework. We also briefly discuss the applicability of this algebraic (as opposed to isometric) formalism in contexts where there is no isometry, even locally. Although we do not provide a definitive framework for non-isometric T-duality, we conclude that the examples considered here suggest this formalism has the potential for wider application than the Buscher construction. We also find algebraic structures familiar from doubled geometry \cite{Hull:2009sg, Hull:2007jy, Plauschinn:2018wbo} arising in this context. Introducing the operators on the torus fibre $\Pi_I(\sigma):=(2\pi P_i(\sigma),X'^i(\sigma))$, which transform as vectors under $O(d,d)$, we consider related fields ${\cal A}_M(\sigma)=U_M{}^I(X)\Pi_{I}(\sigma)$. In the case where the background is a $T^{d-1}$ bundle over $S^1$ with monodromy $e^N\in O(d-1,d-1)$, $U$ acts on the fibres as $e^{-Nx}$, whilst the base is left unchanged. The canonical commutation relations are shown to satisfy an algebra\footnote{The notation we adopt is
\begin{equation}
\delta'(\sigma-\sigma'):=\frac{\rd}{\rd\sigma}\delta(\sigma-\sigma')=-\frac{\rd}{\rd\sigma'}\delta(\sigma-\sigma')=:-\delta'(\sigma'-\sigma).
\end{equation}}
\begin{equation}\label{A}
[{\cal A}_I(\sigma),{\cal A}_J(\sigma')]=2\pi iL_{IJ}\delta'(\sigma-\sigma')-2\pi it_{IJ}{}^K{\cal A}_K(\sigma')\delta(\sigma-\sigma'),
\end{equation}
which is reminiscent of that found in (parallelizable) flux compactifications in supergravity \cite{Hull:2006tp} and in doubled geometry \cite{Hull:2009sg, Hull:2007jy}. The algebra (\ref{A}) appears in \cite{Siegel:1993th} many years before doubled geometry was introduced and similar structures have also been studied in related contexts \cite{Osten:2019ayq, Andriot:2012vb}. In section \ref{s:relationship to doubled formalism} we compare this algebra to that found in the doubled formalism and see that, remarkably,  the full doubled algebra emerges  from the torus bundle construction directly from a statement of the monodromy of the fibration. This is in contrast to the conventional doubled geometry constructions, where the doubled Lie group generally requires explicit dependence on \emph{all} doubled coordinates to describe the doubled algebra. In section \ref{s: beyond torus bundles} we briefly consider generalisations of the doubled algebra found in the torus bundle case; firstly, from torus bundles to more general parallelisable spaces, and then to a local construction that may have applicability to a wider class of target spaces, both geometric and non-geometric. We follow this rather speculative section with a brief discussion of future directions in section \ref{s: discussion}.

\section{An Algebraic Approach To T-Duality}\label{s: an algebraic approach to T-duality}

Most work on T-duality has focused on the Buscher construction, which gives central importance to the existence of compact abelian isometries \cite{Buscher:1987sk}. The observation of \cite{Dine:1989vu}, that T-duality may be thought of as a residual discrete gauge symmetry, provides a framework in which to think about T-duality without reference to isometries of the target space.

In this section we review the approach to realising T-duality as a $\Z_2$ automorphism of the operator algebra of the worldsheet theory. For illustrative purposes we focus on the simplest case where the target space is a circle at the self-dual radius $R=\sqrt{\alpha'}$. In this special case, the $\Z_2$ is a discrete subgroup of the larger $SU(2)_L\times SU(2)_R$ automorphism that appears at the self-dual radius. We shall discuss more general cases in the following section.

\subsection{Universal Coordinates}

We start with the action for a string theory on a torus
\begin{equation}\label{Action}
S = -\frac{1}{4\pi}\int_0^{2\pi}\rd\sigma\int \rd\tau \left(\sqrt{\gamma}\gamma^{\alpha\beta}\partial_\alpha X^i \partial_\beta X^j g_{ij}+\epsilon^{\alpha\beta}\partial_\alpha X^i \partial_\beta X^j B_{ij}\right),
\end{equation}
where we have absorbed all of the $\alpha'$ and physical length scales, such as the radius of the circles of the tori, into the metric and B-field. The embedding fields $X^i$ on the circles are all dimensionless and obey the standardised periodicity relations $X\sim X+2\pi$ and we shall take the worldsheet metric to be the Minkowski metric up to Weyl rescaling. The general mode expansion on the cylinder is
\begin{equation}\label{X}
X^i(\sigma,\tau)=x^i+\omega^i\sigma+\tau g^{ij}(p_j-B_{jk}\omega^k)+\frac{i}{\sqrt{2}}\sum_{n\neq 0}\frac{1}{n}\left(\alpha_n^ie^{-in(\tau-\sigma)}+\bar{\alpha}_n^ie^{-in(\tau+\sigma)}\right),
\end{equation}
and the conjugate momentum $P_i = \frac{1}{2\pi}(g_{ij} \Dot{X}^j+B_{ij}X'^j)$ has the mode expansion
\begin{equation}
\Pi_i(\sigma, \tau) \equiv 2\pi  P_i(\sigma, \tau) = p_i + \frac{1}{\sqrt{2}}\sum_{n\neq 0}\left(E^T_{ij}\alpha_n^j e^{-in(\tau-\sigma)}+E_{ij}\bar{\alpha}_n^j e^{-in(\tau+\sigma)}\right),
\end{equation}
where $E_{ij}=g_{ij}+B_{ij}$ is the background field tensor. At fixed $\tau$ (which we can take to be $\tau=0$) these fields obey the equal time commutation relations 
\begin{equation}
[X^i(\sigma), \Pi_j(\sigma')] = 2\pi i \delta(\sigma-\sigma')\delta\indices{^i_j}.
\end{equation}
We take this relationship and the fields $X^i(\sigma)$ and $\Pi_i(\sigma)$ to be background independent, i.e. objects that we can compare across different backgrounds\footnote{We are really only requiring that these fields are universal in a local region of the space of backgrounds that we are interested in.}. This idea is an old one and is discussed for example in \cite{Kugo:1992md}, where $X^i(\sigma)$ and $\Pi_i(\sigma)$ are described as \emph{universal coordinates}.
For now, we shall restrict attention to trivial torus bundles with constant $B$-field. The space of such backgrounds is given by the orbifold
\begin{equation}\label{moduli}
\mathscr{M}_d= O(d,d;\Z)\backslash O(d,d)/O(d)\times O(d),
\end{equation}
where $d$ is the dimension of the torus and the data $E_{ij}$, modulo symmetries, specifies a point on $\mathscr{M}_d$. The requirement that $X^i(\sigma)$ and $\Pi_i(\sigma)$ remain the same as we change the background $E_{ij}$ means that the oscillator modes must be background dependent. Identifying $X^i(\sigma)$ and $\Pi_i(\sigma)$ on different backgrounds, $E_{ij}$ and $E'_{ij}$, requires that the modes are related by\footnote{This quantisation at $\tau=0$ is related to a quantisation at generic $\tau$ by the usual relations $\alpha_n(\tau)=e^{-i\tau L_0}\alpha_n(0) e^{i\tau L_0}=e^{-in\tau}\alpha_n(0)$, where the second equality follows from standard commutation relations. We notice that $X(\sigma,0)\rightarrow X(\sigma,\tau)$ if we also change $x\rightarrow  x^i+\tau g^{ij}(p_j-B_{jk}w^k) \equiv x^i(\tau)$, which has a clear interpretation as a translation arising from a unitary time evolution, i.e. $X^i(\sigma,\tau)$ takes the same algebraic form as $X^i(\sigma)$, but with the replacements of $x^i(\tau)$ and $\alpha^i_n(\tau)$ for $x^i$ and $\alpha^i_n$. Thus, universal coordinates defined for generic fixed $\tau$ have the same algebraic form as the $\tau=0$ case, so the notion of a universal coordinate does not depend on the choice of $\tau$ (but a choice must be made).}
\begin{equation}\label{alpha}
2g'_{ij}\alpha^j_n(E')=\big(E_{ij}^T+E'_{ij}\big)\alpha^j_n(E)+\big(E_{ij}-E'_{ij}\big)\bar{\alpha}^j_{-n}(E),
\end{equation}
\begin{equation}\label{alphaBar}
2g'_{ij}\bar{\alpha}^j_n(E')=\big(E_{ij}^T-{E'}_{ij}^T\big)\alpha^j_{-n}(E)+\big(E_{ij}+{E'}_{ij}^T\big)\bar{\alpha}^j_{n}(E),
\end{equation}
where $\alpha^i_n(E), \Bar{\alpha}^i_n(E)$ are the oscillator modes of the string embedding into the background specified by $E_{ij}$. The momenta and winding $p_i$ and $\omega^i$ do not change (they are defined as radius independent in our conventions). We can see that the $\alpha_0^i$ and $\bar{\alpha}_0^i$ transform in line with (\ref{alpha}) from their definitions
\begin{equation}
g_{ij}\alpha^j_0=\frac{1}{\sqrt{2}}\Big(p_i-E_{ij}\omega^j\Big),      \qquad  g_{ij}\bar{\alpha}^j_0=\frac{1}{\sqrt{2}}\Big(p_i+E^T_{ij}\omega^j\Big).
\end{equation}
It will be useful to introduce the objects, for fixed $\tau$,
\begin{equation}\label{LR Derivatives}
\partial X_i(\sigma)=\frac{1}{2}\Big(\Pi_i(\sigma)-E_{ij}X'^j(\sigma)\Big),  \qquad
\Bar{\partial}X_i(\sigma)=\frac{1}{2}\Big(\Pi_i(\sigma)+E^T_{ij}X'^j(\sigma)\Big),
\end{equation}
where the $i,j$ indices are lowered by $g_{ij}$. The modes are given by
\begin{equation}
\alpha^i_n(\tau)=\frac{\sqrt{2}}{2\pi}\oint\rd\sigma \,\partial X^i(\sigma,\tau)\,e^{-in\sigma}.
\end{equation}
It will also be useful to split $X$ into left and right parts, $X=X_L+X_R$, where
\begin{align}
    &X_L(\sigma) = \frac{1}{2}(x-\tilde{x}) -\frac{1}{\sqrt{2}}\alpha_0\sigma + \frac{i}{\sqrt{2}}\sum_{n\neq 0}\frac{1}{n}\alpha_ne^{in\sigma}, \\
    &X_R(\sigma) = \frac{1}{2}(x+\tilde{x}) +\frac{1}{\sqrt{2}}\bar{\alpha}_0\sigma + \frac{i}{\sqrt{2}}\sum_{n\neq 0}\frac{1}{n}\bar{\alpha}_ne^{-in\sigma}, 
\end{align}
where $\tilde{x}$ is the zero mode of the T-dual coordinate $-X_L+X_R$.

This notion of universal coordinates is useful in understanding how the $\Z_2\in SU(2)\times SU(2)$ gauge symmetry at the self-dual radius generalises to other backgrounds.

\subsection{The T-duality charge}

We shall look for a charge $Q$ that generates an automorphism ${\cal A}\rightarrow e^{iQ}{\cal A}e^{-iQ}$ which has the required $\Z_2$ effect\footnote{Note that we could equivalently choose the duality to act as $\Bar{\partial}X\rightarrow-\Bar{\partial}X$ and leave $\partial X$ invariant. This is equivalent to swapping $X'$ and $\Pi$ with an extra minus sign inserted, which also preserves the canonical commutation relations.}: $e^{iQ}\partial X_i e^{-iQ} = -\partial X_i $ and $e^{iQ}\bar{\partial} X_ie^{-iQ} =\bar{\partial} X_i $. The fields $X_L(\sigma)$ and $X_R(\sigma)$ do not have to produce a nice operator algebra independently, but the combinations $X_R(\sigma)\pm X_L(\sigma)$ do. The $\Z_2$ automorphism simply exchanges these two linear combinations, giving a sigma model description in both cases.

Working in $d=1$ for now, we define 
\begin{equation}\label{Q}
    Q_\Lambda  =\Lambda\oint d\sigma \cos(2X_L(\sigma)),
\end{equation}
which we note is invariant under the periodicity $X_L \rightarrow X_L + \pi$. The action of $Q_\Lambda$ on $\partial X(\sigma)$ may be written as
\begin{equation}
e^{iQ_\Lambda} \partial X(\sigma)e^{-{iQ_\Lambda}} = \partial X(\sigma) + i[Q_\Lambda, \partial X(\sigma)] + \frac{i^2}{2!}[Q_\Lambda,[Q_\Lambda,\partial X(\sigma)]] + ...,
\end{equation}
with the ellipsis denoting nested commutators at higher order in $\Lambda$. Since we are in one dimension and at the self-dual radius, we don't need to worry about indices being raised or lowered (we take the relevant component of the metric to be normalised to unity). Computing the leading contributions gives\footnote{Note also that
\begin{equation}
    \partial X = \frac{1}{2}(\Pi-X')
    = \frac{1}{2}(-X'_L+X'_R - X'_L-X'_R) = -X'_L.
\end{equation}}: 
\begin{align}
    [Q_\Lambda, \partial X(\sigma)] =& \Lambda\oint d\sigma' [\cos(2X_L(\sigma')), -X'_L(\sigma)] \nonumber\\
    =& -\Lambda\oint d\sigma'\left[1-\frac{2^2}{2!}X_L^2(\sigma')+..., X_L'(\sigma)\right] \nonumber\\
    =& -2\pi \Lambda i\sin(2X_L(\sigma)),
\end{align}
and
\begin{align}
   [Q_{\Lambda},[Q_{\Lambda}, \partial X(\sigma)]] = & - 2\pi\Lambda^2 i \oint \rd\sigma'\,[\cos(2X_L(\sigma)), \sin(2X_L(\sigma'))]\nonumber\\
   =& (2\pi )^2  \Lambda^2 \oint \rd\sigma' \,\delta(\sigma-\sigma')\partial X(\sigma') \nonumber\\
   =& (2 \pi \Lambda)^2\partial X(\sigma).
\end{align}
It is clear that the next term $[Q_{\Lambda},[Q_{\Lambda},[Q_{\Lambda}, \partial X(\sigma)]]]$ is proportional to $\sin(2X_L(\sigma))$. Continuing in this fashion, we see that successive nested commutators will alternately give terms proportional to $\sin(2X_L)$ and $\partial X$ with coefficients that are straightforward to determine. Putting this all together gives \cite{Evans:1995su}
\begin{align}
    \notag&e^{iQ_\Lambda}\partial X(\sigma) e^{-iQ_{\Lambda}} 
    \\\notag=& \left(1+\frac{(2\pi\Lambda i)^2}{2!}+\frac{(2\pi\Lambda i)^4}{4!}+...\right)\partial X(\sigma) - i \left(2\pi\Lambda i + \frac{(2\pi\Lambda i)^3}{3!}+\frac{(2\pi\Lambda i)^5}{5!}+...\right)\sin(2X_L(\sigma))
    \\=& \cos(2\pi\Lambda )\partial X(\sigma)+\sin(2\pi\Lambda )\sin(2X_L(\sigma)).
\end{align}
In our conventions, the choice $\Lambda=1/2$ gives the required transformation and so if we define $Q\equiv Q_{1/2}$, we have
\begin{equation}
    e^{iQ}\partial X(\sigma) e^{-iQ}= -\partial X(\sigma).
\end{equation}
Thus, $Q$ is a suitable T-duality charge. Of course, at the self-dual radius, any choice of $\Lambda$ generates a symmetry of the theory as $Q_{\Lambda}$ is built from a $(1,0)$ current. What is interesting is that, even away from the self-dual radius, the charge $Q$ defined above still gives a symmetry of the theory, as we discuss below. Note that we can equivalently use the charge 
\begin{equation}
        Q  =\frac{1}{2}\oint d\sigma \sin(2X_L(\sigma)).
\end{equation}
We get exactly the same transformation either way.  We shall discuss the relationships between these charges in section \ref{ss: gauge equivalence of T-duality charges}.

\subsection{Automorphisms away from the self-dual radius}\label{ss: automorphisms away from the self-dual radius}

Given that the $\Z_2$ duality described above is, at the self-dual radius, a subgroup of an exact gauge symmetry group of the target space theory, it is hardly surprising that the duality is a symmetry of the theory. What is more surprising is that the $\Z_2$ continues to hold as an exact symmetry of the theory for any radius. In this section we review the arguments that lead to this conclusion, but from the algebraic perspective rather than the usual Buscher construction.

Throughout this paper, the background $E = G$ will refer to metric $G=1$ and B-field $B=0$, i.e. the background at which we have an enhanced $SU(2)$ gauge symmetry. From the derivatives \eqref{LR Derivatives}, we have
\begin{equation}
    \partial X_i  \equiv \partial X_i (G)=\frac{1}{2}(\Pi_i  - G_{i j }X'^j ).
\end{equation}
We can relate $\partial X_i(E)$ and $\Bar{\partial}X_i(E)$ at different backgrounds using their expressions in terms of the background independent fields given above. Rearranging to get 
\begin{equation}
    \Pi_i-B_{ij}X'^j=\frac{1}{2}\Big(\partial X_i+\bar{\partial}X_i\Big), \qquad  g_{ij}X'^j=-\frac{1}{2}\Big(\partial X_i-\bar{\partial}X_i\Big),
\end{equation}
we find that 
\begin{align}
    \label{dXEinTermsofdXE'}\partial X_i(E) &= \frac{1}{2}g'^{jk}\Big((E_{ij}+E'^T_{ij})\partial X_k(E')+(-E_{ij}+E'_{ij})\bar{\partial} X_k(E')\Big), \\
    \label{dbarXEinTermsofdXE'}\bar{\partial} X_i(E) &= \frac{1}{2}g'^{jk}\Big((-E^T_{ij}+E'^T_{ij})\partial X_k(E')+(E^T_{ij}+E'_{ij})\bar{\partial} X_k(E')\Big).
\end{align}
These expressions allow us to determine how composite operators defined in terms of $\partial X_i(E')$ and $\bar{\partial} X_i(E')$ transform under the automorphism generated by charge $Q$ if we know how the operators  $\partial X_i(E)$ and $\bar{\partial} X_i(E)$ transform. So, for example, if $d=1$,
\begin{equation}
    \partial X(E)=\frac{1}{2}G^{-1}\Big((g+G)\partial X+(G-g)\bar{\partial}X\Big),
\end{equation}
from which we deduce
\begin{equation}
   e^{iQ} \partial X(E)e^{-iQ}=\frac{1}{2}G^{-1}\Big(-(g+G)\partial X+(G-g)\bar{\partial}X\Big).
\end{equation}
More generally, if we know how an operator ${\cal F}(E)$ defined at a background $E$ transforms under T-duality, then, if we know the relationship between ${\cal F}(E')$ and ${\cal F}(E)$, we can use the definition of the charge $Q$ in the background $E$ to determine how the symmetry acts on ${\cal F}$ defined at the background $E'$. For example, the chiral stress tensor
\begin{equation}
T(\sigma)=g^{ij}\partial X_i(E)\partial X_j(E)
\end{equation}
is clearly invariant at the self-dual radius, but transforms in a more complicated manner at other points of $\mathscr{M}_d$. It was shown in \cite{Evans:1995su} that the charge of the kind (\ref{Q}) maps a stress tensor to a stress tensor, for generic radius, only for the value of the parameter $\Lambda=1/2$ (in our conventions). Thus, for this value of the parameter, the automorphism is not only a symmetry of the conformal field theory, but relates one stress tensor to an apparently different one, yielding a duality between string theories.

\subsection{Symplectomorphisms and Charge Conservation}

All automorphisms of the operator algebra are symmetries of the theory, but those that preserve the Hamiltonian play a special role. As such, we would like to know under what conditions the charge $Q(\Lambda)=e^{i\Lambda h}$ is conserved. This is important as, if the notion of universal coordinates and canonical commutation relations\footnote{The requirement that the automorphism preserves the commutation relations is already a sign that it must be a symplectomorphism.} is to survive the automorphism, it must make sense at each fixed value of $\tau$. The time evolution of $Q(\Lambda)$ is given by the Hamiltonian
\begin{equation}
Q_{\tau}(\Lambda)=e^{-iH\tau}Q_0(\Lambda)e^{iH\tau},
\end{equation}
or $e^{iQ_{\tau}}=e^{-iH\tau}e^{iQ_0}e^{iH\tau}$. For the charge to be conserved, we require $Q_{\tau}(\Lambda)=Q_0(\Lambda)$, i.e.
\begin{equation}
e^{i\Lambda h}=e^{-iH\tau}e^{i\Lambda h}e^{iH\tau}.
\end{equation}
We can rewrite this as
\begin{equation}
e^{iH\tau}=e^{-i\Lambda h}e^{iH\tau}e^{i\Lambda h},
\end{equation}
which says that the automorphism preserves the Hamiltonian. This can happen in two ways. The most obvious way is if $[H,h]=0$, i.e. the functional $h$ is constant in time. This will be true if we can write $h$ as
\begin{equation}
h=\oint\rd\sigma J(\sigma),
\end{equation}
where $J(\sigma)$ is a weight $(1,0)$ or $(0,1)$ current. In that case, $[H,J(\sigma)]=0$ and the current is conserved. This is the case when $J(\sigma)$ generates a continuous symmetry of the theory, such as $B$-field gauge transformations, spacetime diffeomorphisms, or the $SU(2)\times SU(2)$ gauge symmetry at the self-dual radius. As the symmetry is continuous, $\Lambda$ can take any value in the parameter space of the corresponding Lie group.

This is not the only way to preserve the Hamiltonian. Consider a theory with action (\ref{Action}), taking the worldsheet metric to be $h_{ab}=\text{diag}(1,-1)$. The Hamiltonian may be written as
\begin{equation}\label{HamiltonianInt}
H=\oint\rd\sigma\;{\cal S}^T{\cal H}{\cal S},
\end{equation}
where ${\cal S}_I=(\Pi_i,X'^{i})$, where $\Pi_i$ is the canonical momentum and
\begin{equation}
{\cal H}^{IJ}=\left(\begin{array}{cc}
   g^{ij}   & g^{ik}B_{kj} \\
B_{ik}g^{kj}  & g_{ij} +B_{ik}g^{kl}B_{lj}
\end{array}
\right).
\end{equation}
Indices are raised and lowered by the invariant of $O(d,d)$; ${\cal S}^I=L^{IJ}{\cal S}_J=(X'^{i},\Pi_i)$, where
\begin{equation}
    L^{IJ} = \begin{pmatrix}0 & 1\\1 & 0\end{pmatrix}.
\end{equation}
The Hamiltonian is invariant under the transformations
\begin{equation}
{\cal S}\rightarrow {\cal O}{\cal S},   \qquad  {\cal H}\rightarrow {\cal O}{\cal H}{\cal O}^{-1},
\end{equation}
where ${\cal O}\in O(d,d;\Z)$ (the discreteness is required to preserve integer-valuedness of the zero modes). Thus, an automorphism that generates an $O(d,d;\Z)$ transformation,
\begin{equation}
e^{-iQ}{\cal S}_Ie^{iQ}={\cal O}_I{}^J{\cal S}_J,
\end{equation}
will also preserve the Hamiltonian (provided ${\cal H}^{IJ}$ is also transformed). The requirement that ${\cal O}\in O(d,d;\Z)$ means that $\Lambda$ may only take certain discrete values. Thus, we see that the condition for charge conservation is that the charge generates a symplectomorphism.

\subsection{Gauge equivalence of T-duality charges} \label{ss: gauge equivalence of T-duality charges}

In this subsection, it will be more convenient to use OPEs instead of commutation relations. Therefore, the charge is now\footnote{The superscript in $Q^c$ refers to the cosine function that is used to define the charge. It is included here to distinguish it from a similar construction using the sine function.}
 \begin{equation}
 Q^c=  \frac{1}{2}\oint \rd z \cos(2X_L(z)).
 \end{equation}
We also use conventions where the $XX$ OPE is given by
\begin{equation}\label{XXOPE}
    X(z)X(w)\sim -\pi \log|z-w|^2
\end{equation}
at the self-dual radius, which is the only radius we will be interested in for this subsection. To show gauge equivalence of the charges, we need to show that the transformations they induce are the same up to a $U(1)_L\times U(1)_R$ gauge transformation. It is sufficient to look at general derivatives of $X$ and general exponentials of $X$\footnote{This follows from the fact that
\begin{equation}
    e^{iQ}ABe^{-iQ} = e^{iQ}Ae^{-iQ}e^{iQ}Be^{-iQ}
\end{equation}
for any operators $A,B$, where we rely on the associativity of the OPE.}. The transformation of higher derivatives of $X$ under an automorphism follows straightforwardly once the transformation of $\partial X$ is known,
\begin{equation}\label{higherDerivativeTransformation}
    e^{iQ}\partial^n X e^{-iQ} = \partial^{n-1} e^{iQ}\partial X e^{-iQ} = -\partial^n X,
\end{equation}
and this is the same for both charges. However, the transformation of exponentials $e^{inX}$ is more difficult. In \cite{Evans:1995su} the transformation of such exponentials is found using a point-splitting argument, but we present a slightly different approach, via induction, and the details are presented in Appendix \ref{a: vertex operators and WZW}. The result of the automorphsim is
\begin{equation}\label{expGeneraln1}
    e^{inX_L(z)} \rightarrow \left\{
    \begin{matrix}
        ie^{-inX_L(z)}, \qquad \text{n odd} \\
        e^{-inX_L(z)}, \qquad \text{n even}.
    \end{matrix}\right.
\end{equation}
We see that, in the $n$ odd case, there is an extra factor of $i$ compared to expectations. This technical detail is discussed in Appendix \ref{a: vertex operators and WZW}.

Thus far we have made use of the $\Z_2$ symmetry generated by the charge (\ref{Q}) with $\Lambda=1/2$; however, we could have equally well used the charge
\begin{equation}
Q^s= \frac{1}{2}\oint \rd z \sin(2X_L(z)).
\end{equation}
Indeed, it was this charge that was used in \cite{Evans:1995su}. Both charges give rise to the same action $\partial X \rightarrow -\partial X$, but they do not act in the same way on exponentials. This may seem strange at first, but it is not hard to see that these charges are related by a $U(1)_L\times U(1)_R\subset SU(2)_L\times SU(2)_R$ gauge transformation generated by the currents $\partial X(z)$ and $\bar{\partial}X(\bar{z})$. Moving away from the self-dual radius, this symmetry is preserved, so this equivalence of the charges holds throughout moduli space\footnote{The case where $U(1)_L\times U(1)_R$ is also broken will be discussed elsewhere \cite{Mahmood202?}.}. If we use the sine charge, via a similar process to the above, we find that 
\begin{equation}
    e^{inX_L} \rightarrow \left\{
    \begin{matrix}
        ie^{in(-X_L+\pi/2)}, \qquad \text{n odd} \\
        e^{in(-X_L+\pi/2)}, \qquad \text{n even},
    \end{matrix}\right.
\end{equation}
where it is instructive to write an $n$-dependent phase on the right hand side as a shift in $X_L$. Written in this way, it seems that the effects of the two charges on $e^{inX_L}$ are related by a $U(1)_L\times U(1)_R$ transformation that gives the required shift in $X_L$. To see this, consider the automorphism generated by the charge
\begin{equation}
    Q^3_\Lambda \equiv \Lambda \oint\rd z \partial X(z), 
\end{equation}
where the 3 superscript indicates that it is the charge corresponding to the current $J^3$. By a simple OPE calculation, we can show that 
\begin{equation}
    e^{iQ^3_\Lambda}e^{inX_L}e^{-iQ^3_\Lambda} = e^{-in\pi\Lambda}e^{inX_L},
\end{equation}
or simply $e^{iQ^3_\Lambda}X_Le^{-iQ^3_\Lambda} = X_L - \pi\Lambda$. Now, denote the T-duality cosine and sine charges by $Q^c, Q^s$ respectively. 

If we set $\Lambda=-1/2$ in $Q^3_\Lambda$, we find that $e^{inX_L} \rightarrow (-1)^{\frac{n}{2}}e^{inX_L}$. Thus, we have 
\begin{equation}
    e^{inX_L}\xrightarrow{Q^s} \left\{ 
        \begin{matrix}
            ie^{in(-X_L+\pi/2)} \\
            e^{in(-X_L+\pi/2)} 
        \end{matrix}\right.
        \xrightarrow{Q^3_{-\frac{1}{2}}} \left\{ 
        \begin{matrix}
            ie^{-inX_L}, \qquad \text{n odd}\\
            e^{-inX_L}, \qquad  \quad \text{n even}.
        \end{matrix}\right.
\end{equation}
The right-hand side is the same as the $Q^c$ transformation, and thus we have shown that the effects of the two charges are related by $U(1)_L$ gauge transformations. The difference between the cases for $n$ odd or even can be traced to the way in which the highest weight states in the modules $L_{[1,0]}$ and $L_{[0,1]}$ transform under the automorphisms generated by $Q^{c,s}$ (see the discussion in Appendix \ref{a: vertex operators and WZW}). It is not hard to show that the charges $Q^c$ and $Q^s$ themselves are directly related by a similar $U(1)$ gauge transformation, generated by $Q^3_\Lambda$ with parameter $\Lambda=-1/4$, i.e. $UQ^cU^{-1}=Q^s$, where $U=\exp\left({-\frac{1}{4}\oint}\rd z\,J^3(z)\right)$.

\section{Torus Bundles}
\label{s: torus bundles}

In the previous section, we have illustrated how, using knowledge of how the operator algebra defined at a point of enhanced symmetry transforms under the automorphism generated by $Q$, one can determine  the effect of the automorphism at a general radius.

\cite{Evans:1995su, Evans:1989cs, Evans:1989hh} also consider torus backgrounds with constant $B$-field. These are exact string backgrounds and have an explicit description in terms of a worldsheet CFT. However, the realisation of the approach to T-duality proposed in \cite{Evans:1995su} is contingent on being able to describe the Hilbert space of the CFT corresponding to a particular background in terms of the Hilbert space of the theory at the self-dual point.

We are interested in generalising this construction and in this section we consider two key issues. The first is the question of what the appropriate framework to discuss and generalise this construction outlined in section \ref{ss: automorphisms away from the self-dual radius} is. We shall argue that the identification of a connection on the space of backgrounds with which parallel transport may be performed (from a point of enhanced symmetry to a background of interest) provides a natural generalization of the technique of section \ref{ss: automorphisms away from the self-dual radius}. A second issue is to identify a class of backgrounds suitable for practical application of these techniques. Torus bundles with monodromy provide a useful set of examples. A third question, that of what obstructions there are to further generalisation, is touched upon but will be largely left open.

\subsection{Connections on the space of backgrounds}

Given a connection $\Gamma$ on the space of backgrounds, the procedure of section \ref{ss: automorphisms away from the self-dual radius} can be realised by using the connection to parallel transport the fields from one point on the space of backgrounds to another. The key observation of \cite{Ranganathan:1992nb} is that the universal coordinate argument \cite{Kugo:1992md} that leads to (\ref{dXEinTermsofdXE'}) and (\ref{dbarXEinTermsofdXE'}) can be understood infinitesimally\footnote{There can be obstructions to integrating up some connections to finite changes of background \cite{Ranganathan:1993vj}.} in terms of parallel transport using a connection on the space of toroidal backgrounds (\ref{moduli}). Connections on the space of string backgrounds have been explored in \cite{Ranganathan:1992nb,Ranganathan:1993vj}, where a very general class of CFT connections was considered. The choice of which connection to use depends on what we want to preserve under parallel transport.

\subsubsection{CFT connections and surface states}

Given that a string background (in the perhaps limited sense of satisfying the string equations) is equivalent to the existence of an associated worldsheet CFT, natural things to preserve under parallel transport are gluing properties of CFTs \cite{Ranganathan:1993vj}. In this way, the parallel transport of the state space of a CFT will yield another CFT and in principle could be used to explore the CFT moduli space.

A perturbative string theory defined on a certain background, in the form of a string field theory, may be described by the BRST charge ${\cal Q}_B$ and the collection of surface states $\langle \Sigma_n|$. As such, a simple way to understand  the effect of a connection is to see how the stress tensor and surface states transform under parallel transport. Let $\Sigma_{n,g}$ be an $n$-pointed Riemann surface of genus $g$. A surface state $\langle \Sigma_{n,g}|$ may be defined as follows. Given the states $|\Psi_i\rangle$ in the string Hilbert space at the $i$'th puncture, the correlation functions $\langle \Psi_i(z_1)...\Psi_n(z_N)\rangle$ are given by\footnote{
The surface state for the matter sector may be written as 
\begin{equation}
\big\langle \Sigma_n\big|=\big\langle\vec{0}\big|\exp\left(\frac{1}{2}\sum_{a,b=1}^N\sum_{m,p\geq 0}{\cal N}_{mp}(z_a,z_b)\alpha^{(a)}_m\cdot \alpha_p^{(b)}+\text{a.h.}\right),
\end{equation}
where ${\cal N}_{mp}(z_a,z_b)$ are Neumann coefficients (see, for example \cite{Kugo:1989tk}) and 
$\big\langle\vec{0}\big|=\langle 0_1|...\langle 0_n|(2\pi)^D\delta^D\left(\sum_ap_a\right),
$ and similarly for the right-moving sector. Ghost parts have been neglected but may be straightforwardly incorporated.}
\begin{equation}
\langle \Psi_i(z_1)...\Psi_n(z_n)\rangle=\langle \Sigma_n||\Psi_1\rangle ...|\Psi_n\rangle,
\end{equation}
where $|\Psi_i\rangle$ is the state corresponding to the operator $\Psi_i$ inserted at the puncture located at $(z_i,\bar{z}_i)$.

If we have a CFT, the general framework for constructing connections on the space of backgrounds has been mapped out \cite{Campbell:1990dz}  (see also \cite{Ranganathan:1993vj}) in terms of the surface state $\big|\Sigma, z_i\big\rangle$. Such connections give a manifestly conformally-invariant way of moving between backgrounds. Moving from the point $s^{\alpha }$ to $s^{\alpha }+\delta s^{\alpha}$ on the space of backgrounds may be achieved in a CFT by using a marginal operator $\Phi_{i }$, conjugate to the deformation, giving the connection
\begin{equation}
    \big|\Sigma_N\big\rangle_{E'}=\big|\Sigma_N\big\rangle_E+\delta s^{\alpha }{\cal D}_{\alpha }\big|\Sigma_N\big\rangle_E+...,
    \end{equation}
where ${\cal D}_{\alpha}=\partial_{\alpha}+\Gamma_{\alpha}$. The connection $\Gamma_{\alpha}$ is given by \cite{Ranganathan:1993vj, Campbell:1990dz}
    \begin{equation}\label{connection}
    {\cal D}_{\alpha }|\Sigma_N\big\rangle_E=-\int_{\Sigma\backslash \cup_{\alpha} D_{\alpha}}\rd^2z\langle \Phi_{i }|\Sigma_{N+1,g};z\rangle_E-\sum_{a=1}^N\Omega_{\alpha }^{(a)}|\Sigma_{N,g}\rangle_E,
    \end{equation}
    where $|\Sigma_{N+1,g};z\rangle$ is a surface state with an extra puncture at the point $z$ where the marginal operator  $\Phi_{i }$ is inserted. This connection will depend on the choice of $\{D_{\alpha}\}$ cut out of the Riemann surface. The choice of $\{D_{\alpha}\}$ is not the only choice one can make in defining a connection. We can also include an automorphism of the Hilbert space at a point. This is denoted by $\Omega_{\alpha}^{(a)}$ above\footnote{In this framework, one can see how the stress tensor transforms in such a way as to preserve conformal invariance \cite{Campbell:1990dz}. For the stress tensor, we can get the Virasoro generators from the two-punctured sphere
\begin{equation}
L_n=\frac{\rd}{\rd \epsilon_n}\big\langle \Sigma; z',1/z'\big|_{\epsilon_n=0},
\end{equation}
where $z'=z+\sum_n\epsilon_nz^n$. The change in the modes of the stress tensor under a change in background may be extracted by
\begin{equation}
\epsilon\delta L_n=\frac{1}{2\pi i}\int_{D'\backslash D}\rd^2z\Phi_{z\bar{z}},
\end{equation}
where the discs $D'$ and $D$ are related by a conformal transformation generated by the stress tensor and $\Phi$ is a marginal operator relating the two backgrounds under consideration.}. Together, $\{D_{\alpha},\Omega^{a}_{\alpha}\}$ specify the connection. Moreover, the choices of $D_{\alpha}$ and $\Omega_{\alpha}$ are not independent and it was shown in \cite{Ranganathan:1993vj} that a change in one can be given by a change in the other. The utility in introducing an automorphism $\Omega_{\alpha}$ is that it may be used to remove divergences that may appear in the integral term on the right-hand side of (\ref{connection}).

 There are a number of possible CFT connections and the effect of parallel transport by such connections on the surface states provides a useful language in which to frame the issue of background independence in string field theory \cite{Sen:1993mh}. However, it is difficult to find suitably explicit, yet interesting, string solutions which are non-trivial torus fibrations and for which the explicit CFT is known. One can learn a lot by studying explicit toy models which, though not CFT descriptions of complete string backgrounds, may play an interesting role as building blocks for such backgrounds. Well studied examples include $T^2$ bundles with monodromies in the $SL(2;\Z)$ modular group of the fibre and $T^3$ backgrounds with constant volume-filling NS flux. Though not exact string solutions, such examples have been recently studied \cite{Chaemjumrus:2019ipx} and do play a role in what are thought to be string solutions. They also play a role in well-known NS-5 brane and Kaluza-Klein monopole backgrounds \cite{Gregory:1997te, Tong:2002rq, Harvey:2005ab, Jensen:2011jna}. Written in terms of surface states, the CFT connection (\ref{connection}) is not obviously applicable to more general sigma models. However, the formalism may be adapted to more general quantum field theories.

\subsubsection{A connection from Universal Coordinates}

There are therefore a large number of possible CFT connections, specified by the choice $\{D_{\alpha},\Omega_{\alpha}\}$ and a marginal operator $\Phi_{i }$. Which connection we choose depends on what aspects of the model we care about. It was shown in \cite{Sen:1993mh} that String Field Theory selects, as a natural connection, that found in \cite{Kugo:1992md}. In this case, the connection arises not from the principle of gluing being a CFT operation, but from the requirement that the universal coordinates $(\Pi_i(\sigma),X^i(\sigma))$ are preserved on backgrounds of the form $M_D\times T^d$. Put another way, at each point on the space of backgrounds there exists a canonical set of fields $(\Pi_i(\sigma),X^i(\sigma))$. This connection transports the $(\Pi_i(\sigma),X^i(\sigma))$ at one background to the corresponding $(\Pi_i(\sigma),X^i(\sigma))$ at another background, where they are identified (up to possible symmetry transformations at that point). This connection was originally studied in \cite{Ranganathan:1993vj} as an example of a CFT connection for which the discs excised from the worldsheet in (\ref{connection}) are unit discs and there is no $\Omega$ term. This is the simplest choice of connection intuitively from the perspective of string field theory, but there are technical challenges to integrating infinitesimal deformations up to finite changes of background.

It was shown in \cite{Ranganathan:1992nb} that this CFT connection is equivalent to a connection proposed in \cite{Kugo:1992md} when considering string field theory on toroidal backgrounds. The virtue of the derivation of this connection given in \cite{Kugo:1992md} is that it does not rely on CFT concepts and so generalises to sigma models that are not exact CFTs. In this sense, it is a connection that can allow us to study backgrounds that are, from a string field theory perspective, off-shell, opening up the possibility of connecting a very wide class of sigma models to one another by parallel transport of this connection. 

To start with, let us stick with CFTs. The basic idea is to consider an object, such as the surface state defined at a background $E$, and then to consider how it changes under a change of background. To first order, this gives the connection
\begin{equation}
|\Sigma_n\rangle_{E+\delta E}-|\Sigma_n
\rangle_E=\delta s^{\alpha}\Gamma_{\alpha}|\Sigma_n\rangle+...,
\end{equation}
where $\delta s^{\alpha}$ is an infinitesimal motion in the space of backgrounds. As discussed in section \ref{s: an algebraic approach to T-duality}, $\partial X_i(E)$ defined at the background $E_{ij}$ is related to that defined at the background with metric $G_{ij}$ and zero $B$-field by
\begin{equation}\label{1}
\partial X_i(E)=\frac{1}{2}(\partial X_i+\bar{\partial} X_i)+\frac{1}{2}E_{ij}G^{jk}(\partial X_k-\bar{\partial} X_k),
\end{equation}
where the $\partial X_i$ on the right-hand side of (\ref{1}) are defined at the background with metric $G_{ij}$. Given that the metric and $B$-field are constant, the associated mode transformation is given by (\ref{alpha}). In terms of the circle at radius $R=\sqrt{\alpha'}+\delta R$, this latter result gives
    \begin{equation}\label{m}
    \delta\alpha_n=\frac{\delta R}{\sqrt{\alpha'}}\Big(\alpha_n+\bar{\alpha}_{-n}\Big),
    \end{equation}
to leading order in $\delta R$, as found in \cite{Kugo:1992md}. The surface state $|\Sigma_n\rangle$ depends explicitly on these modes and so might be expected to transform under the change of background. The Neumann coefficients do not change, but the $\alpha_n$ do, and with the change (\ref{m}) comes a corresponding change in the worldsheet vacuum. The vacuum is defined by $\alpha_n|0\rangle=0=\bar{\alpha}_n|0\rangle$ for $n\geq0$, and so a change in background also changes the vacuum,
    \begin{equation}
    |0\rangle\rightarrow e^{\Delta}|0\rangle,  \qquad  \Delta=\sum_{n\neq 0}\frac{1}{2n}\alpha_n^i\delta E_{ij}\bar{\alpha}_n^j.
    \end{equation}
   Using these results, it follows \cite{Kugo:1992md, Sen:1993mh} that the $\Omega$-independent part of this connection can be seen to preserve the surface states (for $N>2$). This can be understood from the perspective of the CFT connection as that connection which takes $\{D_{\alpha}\}$ to be unit discs in terms of projective coordinates on $\Sigma_{N,g}$. The sewing relation $w=1/z$ then tells us that the domain of integration $\Sigma\backslash \cup_{\alpha}D_{\alpha}$ in (\ref{connection}) is empty and the connection is of the form
\begin{equation}\label{KZ}
\Gamma_{\alpha }|\Sigma_N\big\rangle_E=\sum_{a=1}^N\Omega_{\alpha }^{(a)}(s)|\Sigma_N\rangle_E,
\end{equation}
i.e. just a point-wise automorphism on the Hilbert space.

An interesting feature of these connections is that they generalise to arbitrary deformations $\delta E_{ij}$ and so we can discuss general nonlinear sigma models, not just CFTs. A difference is that we no longer have a clear notion of what we mean by a surface state $|\Sigma\rangle$ if we are not working exclusively with CFTs; however, this is really just a difference in approach, rather than principle. Connections, such as those discussed above, have been studied for quantum field theories that are not CFTs. In particular, how correlation functions in these theories change as we move from one point in parameter space to another has been studied. We shall say a little more about this in the next section and more details may be found in \cite{Sonoda:1991mv,Sonoda:1992hd,Sonoda:1993dh,Papadodimas:2009eu} and references to be found therein.

Thus, a connection such as that  defined by universal coordinates gives a way to clearly describe the transport of a Hilbert space from one point on the space of backgrounds to another. The Buscher procedure also does not require the worldsheet theory to be a CFT and so this allows us to make contact with the predictions of T-duality applied to more general non-linear sigma models. This observation will form the basis of the application of the procedure of \cite{Evans:1995su} to various toy models based on backgrounds such as the nilfold, $T^3$ with constant $H$-flux and their non-geometric relatives, amongst others.

\subsection{Connections and Torus Bundles}

We consider the role CFT connections and their generalizations can play in understanding a special class of torus bundle backgrounds.

\subsubsection{Twisting as a relationship between backgrounds} \label{ss: twisting as a relationship between backgrounds}

Consider string theory on a $T^2$ background with constant metric $g_{ij}$ and $B$-field $B_{ij}$. The space of such backgrounds is the orbifold
\begin{equation}
\mathscr{M}=O(2,2;\Z)\backslash O(2,2)/O(2)\times O(2),
\end{equation}
and we introduce local coordinates $m^{\alpha}$ on $\mathscr{M}$. Consider a curve $\gamma\subset \mathscr{M}$ with real parameter $s$. Given a connection $\Gamma$, we can construct a covariant derivative ${\cal D}_{\alpha}=\partial_{\alpha}+\Gamma_{\alpha}$ and parallel transport elements of the Hilbert space $|\psi\rangle$ according to ${\cal D}\psi=0$. Thus, we can integrate this expression along a path, parameterised by $s$, to relate operators of the CFT at one point in $\mathscr{M}$ to another: 
\begin{equation}\label{transformation}
\psi^I(s_2)=\exp\left(-\int_{s_1}^{s_2}\rd s\frac{\rd m^{\alpha}}{\rd s}\Gamma_{\alpha J}^I\right)\psi^J(s_1).
\end{equation}
Consider a path $\gamma$ such that any two points on $\gamma$ are related by an action of a particular generator of $O(2,2)$. For simplicity, we take the parameterisation to be aligned with one of the local coordinates $m$ on $\mathscr{M}$ and also $(s_1,s_2)=(0,m)$. The mode operators $(\alpha_n,\bar{\alpha}_n)$ provide illustrative examples of operators we might transport from one point to another. Alternatively, we could take $\psi^I$ to include the worldsheet currents $\partial X^a$ and $\bar{\partial}X^a$. Suppose we start at a background with zero $B$-field and metric $G_{ab}$. In general the transformation of a basis of operators given by such a parallel transport will mix the fields together. Matters are simpler for the deformations that move us around the space of torus compactifications. The transformation of the modes is given by \eqref{alpha}, where we note that the zero modes transform amongst themselves and so the transformations of the fields $\partial X_a$ and $\bar{\partial}X_a$ tell us exactly how the target space changes under the transformation and we shall focus on those fields. For different deformations of the theory, different subsets of fields may be of interest\footnote{This would be the case if, for example, some additional fields became massless along the path.}. The relationship (\ref{transformation}) can be written as
\begin{equation}
\widehat{\cal A}_I=\Big(e^{-\Gamma(m)}\Big)_I{}^J{\cal A}_J,
\end{equation}
where we have written ${\cal A}_I=(\partial X_a,\bar{\partial}X_b )$. In particular,
\begin{equation}\label{me}
\partial \widehat{X}_a=\frac{1}{2}\Big(\delta_a^b+E_{ac}G^{cb}\Big)\partial X_b+\frac{1}{2}\Big(\delta_a^b-E^T_{ac}G^{cb}\Big)\bar{\partial}X_b,
\end{equation}
where $(\partial \widehat{X}_a,\bar{\partial} \widehat{X}_a)$ is defined at the background $E_{ab}$ and $(\partial X_a,\bar{\partial}X_a)$ is defined at the background $G$. For example, we could take $\partial X_a=(\partial Y,\partial Z)$ and the deformation to generate a constant $B$-field
\begin{equation}
B=m\rd Y\wedge \rd Z,
\end{equation}
giving
\begin{equation}
\partial\widehat{Y}=\partial Y+\frac{m}{4}(\partial Z-\bar{\partial}Z),	\qquad	\partial\widehat{Z}=\partial Z-\frac{m}{4}(\partial Y-\bar{\partial}Y).
\end{equation}
Such a deformation can be written as an automorphism
\begin{equation}\label{sim}
\widehat{\cal A}_I=\exp\Big(-\Gamma_I{}^J\Big){\cal A}_J=U^{-1}{\cal A}_IU,
\end{equation}
where
\begin{equation}
U=\exp\left(\frac{m}{4}\oint\rd\sigma \Big(Y(\sigma) Z'({\sigma})-Y'(\sigma) Z({\sigma})\Big)\right).
\end{equation}
Since $\widehat{\cal A}_I$ can be written in this way, such a deformation is pure gauge, provided $m$ is appropriately quantised\footnote{Setting $m\in\Z$ sets $e^{\Gamma(m)}\in O(2,2;\Z)$.}.

A second example is the deformation
\begin{equation}
U=\exp\left(-\frac{m}{2}\oint\rd\sigma \Big( Z({\sigma}) \Pi_y(\sigma) \Big)\right).
\end{equation}
This too is a gauge transformation and may be understood as follows. In terms of the action of $\Gamma$ along $\gamma$, we might consider the element of the parabolic conjugacy class of $SL(2)$
\begin{equation}\label{twist}
\Pi_a\rightarrow \big(e^{-\Gamma }\big)_a{}^b\Pi_b,    \qquad X'^a\rightarrow \big(e^{-\Gamma^T }\big)^a{}_bX'^b, \qquad  \Gamma=\left(\begin{array}{cc}
0 & m\\
0 & 0
\end{array}
\right),    \qquad  m\in \R,
\end{equation}
where $\Gamma^T$ denotes the transpose of $\Gamma$ and $e^{\Gamma}\in SL(2)$. Thus the algebra of operators of the theory at the background at one point on $\gamma$ is related to that at another point by an $SL(2)$ transformation. In other words, as we move along the curve $\gamma$ we change the complex structure of the torus. In this way, we can think of $\gamma$ as a curve generated by a particular element of the Lie algebra of $SL(2)$. In particular, if we chose the parameterisation such that the metric at $m=0$ is given by $g_{ab}=G_{ab}$, then the metric at a point $t\neq 0$ would be given by $g(m)=e^{-\Gamma m}G e^{-\Gamma^T m}$. Alternatively, we can describe this by a change in the complex structure,
\begin{equation}
\tau(t)= i+m,
\end{equation}
on the $T^2$ fibre, and by a general element of the Mobius group for a $\gamma$ generated by a general element of $SL(2)$. Any two backgrounds related by $SL(2;\Z)\subset O(2,2;\Z)$ are identified in $\mathscr{M}$ and so are equivalent. This is just the action of the modular group of the $T^2$ in this simple case and we can consider transporting around closed loops, provided the monodromy is in $O(2,2;\Z)$.

\subsubsection{Twisted Backgrounds}

Where things become interesting is when the above discussion inspires constructions of non-trivial backgrounds that may be part of exact string solutions. In this case, we take the closed curve $\gamma$ as a physical direction in spacetime with coordinate $x$ and fibre the $T^2$ over the line, varying the complex structure in the above way as we do so. Moreover, $x$ can be made periodic as long as the resulting monodromy is an element of $O(2,2;\Z)$.

Thus, we have a locally smooth geometry given by a $T^2$ fibred over a circle with coordinate $x$. The monodromy of the bundle is an element of $O(2,2;\Z)$ acting on the fields $\Pi_I:=(\Pi_a,X'^a)$
of the $T^2$ fibre, and here the monodromy along the base direction $x$ must be an element of $O(2,2;\Z)$.

We can use the CFT connection to transport the theory in the fibres from one point to another on the base and, if the fibres contain circles that are at the self-dual radius at some point, then T-duality may be performed as outlined in section \ref{ss: automorphisms away from the self-dual radius} following the procedure of \cite{Evans:1995su}. Treating the CFT in the fibres separately from the base gives a construction in which the duality is performed fibrewise. Such a construction is useful, but may not always give the full story as the base coordinate may play an important role.

The theory in the fibres is a CFT, but by including the base direction with non-trivial monodromy, the background described by the bundle is not a CFT\footnote{Though it may be an important part of an exact string background\cite{Chaemjumrus:2019ipx}.} and so (in the absence of the state operator correspondence) the notion of a surface state is not clear. Instead, we can think about how correlation functions change under deformations of the theory and define the connection correspondingly. The surface state may be expanded in a set of fields $\phi_i$, where $i$ denotes members of the set $S$. The surface state may then be written as
\begin{equation}
\langle\Sigma|=\sum_{i_1,..,i_n}\langle\phi_{i_1}|...\langle\phi_{i_n}|\Sigma_{i_1...i_n},
\end{equation}
where the coefficients $\Sigma_{i_1...i_n}$ are the correlation functions $\Sigma_{i_1...i_n}=\langle \phi_{i_1}...\phi_{i_n}\rangle$. We can therefore couch the definition of the connection in the language of correlation functions as
\begin{equation}
{\cal D}_{\alpha}\Big\langle \prod_{i\in S}\phi_i(z_i)\Big\rangle=-\int_{\Sigma\backslash \cup_i D_{\varepsilon}^{(i)}}\rd^2z\Big\langle {\cal O}_{\alpha}(z)\prod_{i\in S}\phi_i(z_i)\Big\rangle-\sum_{i\in S}\sum_k\Omega_{\alpha i}{}^k\Big\langle \phi_k(z_i)\prod_{j\in S,j\neq i}\phi_j(z_j)\Big\rangle,
\end{equation}
where ${\cal O}_{\alpha}$ is the operator conjugate to the deformation. Indeed, this is closer to the original context in which such connections were first considered in four-dimensional $\phi^4$ and Yang-Mills theories\footnote{The general form of the correlation function for a theory in $D$ dimensions is
\begin{equation}
{\cal D}_{\alpha}\Big\langle \prod_{i\in S} \phi_i(r_i)\Big\rangle=\lim_{\varepsilon\rightarrow 0}\Bigg(-\int_{\Sigma\backslash D_{\varepsilon}} \rd^Dr\Big\langle \widehat{\cal O}_{\alpha}(r) \prod_{i\in S}\phi_i(r_i)\Big\rangle
-\int_{D_1\backslash D_{\varepsilon}}\frac{\rd^Dr}{\text{Vol($S^{D-1}$)}}C_{\alpha i}{}^k\Big\langle \phi_k(r_i)\prod_{j\in S,j\neq i}\phi_j(z_j)\Big\rangle\Bigg).
\end{equation}
There are some differences in higher dimensions, such as the appearance of $\widehat{\cal O}_{\mu}(r)={\cal O}_{\mu}(r)-\langle{\cal O}_{\mu}(r)\rangle$ in the correlation function and the smearing over angular directions. For details, we encourage the reader to consult \cite{Sonoda:1991mv,Sonoda:1992hd,Sonoda:1993dh}. The limit given here reflects the fact that the connection was defined in terms of balls of radius $\varepsilon$ whose size was taken to zero. The prescription chosen to absorb divergences in the integral as $\varepsilon\rightarrow 0$ into $C_{\alpha i}{}^k$ is part of what specifies a choice of connection. The prescription of \cite{Kugo:1992md} excised \emph{unit} discs $D_1$ from the worldsheet and hence exhibits no divergences in the correlation function corresponding to the limit $\varepsilon\rightarrow 0$. As such, $\Omega_{\alpha}$ may be taken to vanish, although one could choose $\Omega_{\alpha}$ to include a finite transformation as discussed above.}. This connection gives the infinitesimal transformation of the fields in the correlation function. The analogue of the operator $e^{-\int\Gamma}$ in (\ref{transformation}) is given by repeated application of the `conjugate' operator ${\cal O}_{\mu}$ (appropriately regularised) and describes a mixing of the fields of the theory as we transport over the base.

We shall apply this general formalism to the torus bundles described above. In what follows, we shall focus only on the $(\partial X_a,\bar{\partial}X_a)$ sector and ignore any possible mixing with other fields. The rationale for this is that, unless modes become massless under parallel transport, such mixing is not expected to play a central role in the description of the target space of the backgrounds we consider to leading order. Neglecting such mixing is equivalent to considering the duality to be performed fibrewise\footnote{Taking the base circle to have radius $R$ and re-introducing the $R$ and $\alpha'$ dependence explicitly, the field $X$ in our expressions is replaced by $X/R$, which may be written as
$$
R^{-1}X(z,\bar{z})=R^{-1}x-iw\ln(z/\bar{z})-i\lambda^2n\ln|z|+\frac{i}{\sqrt{2}}\lambda\sum_{n\neq 0}\frac{1}{n}\Big(\alpha_nz^{-n}+\tilde{\alpha}_n\bar{z}^{-n}\Big),
$$
where $\lambda=\sqrt{\alpha'/R^2}$ and $n$ and $w$ are momentum and winding numbers respectively. The fibrewise construction, in which the field $X$ is taken as a parameter $x$, is recovered in the $w=0$ sector by taking the $\lambda\rightarrow 0$ limit. Thus, we see that the operator mixing that signals a departure from the fibrewise construction enters, at least in the $w=0$ sector, when the supergravity approximation ($\lambda \ll 1$) can no longer be trusted. This is reminiscent of the adiabatic argument of \cite{Vafa:1995gm}.}; however, we hope to discuss such operator mixing in detail elsewhere.

An example of such a construction is the Nilfold, a torus fibred over a circle with monodromy given by \eqref{twist}. In this case, the analogue of  (\ref{me}) is given by
\begin{equation}
\begin{gathered}
\partial\widehat{X}=\partial X, \qquad  \partial\widehat{Y}=\partial Y-\frac{1}{2}mX(\partial Z-\bar{\partial} Z),\\
\partial\widehat{Z}=\left(1+\frac{1}{2}(mX)^2\right)\partial Z-\frac{1}{2}(mX)^2\bar{\partial} Z-\frac{1}{2}mX(\partial Y-\bar{\partial} Y),
\end{gathered}
\end{equation}
and the corresponding expressions for $\bar{\partial} \widehat{X}_a$ given in (\ref{nilfoldAntihol}). One can show that it is not possible to find a $U$ such that this transformation can be produced by a similarity transformation of the form (\ref{sim}) and so this can be thought of as a physical deformation, rather than a gauge transformation. The key difference between the two cases is that here the presence of $X(\sigma)$ multiplying $m$ means that the exponent in $U$ would have to depend linearly on $mX(\sigma)$; however, such a dependence is incompatible with $\partial\widehat{X}=\partial X$. By contrast, in the pure gauge case, $m$ is a parameter - a real number - and so its presence in the exponent of $U$ does not affect the qualitative transformation properties of the fields under automorphisms with $U$.

The above discussion is a rather tortuous way of thinking about familiar duality twist backgrounds \cite{Dabholkar:2002sy,Dabholkar:2005ve}. What is gained by framing the construction in this way is an interpretation of the monodromy as a map between different backgrounds. In particular, we can think of the duality twist (\ref{twist}) as a map from a way of describing a given background $E_{ab}$ in terms of a reference background $G_{ab}$. If we take the reference metric to be at a point of enhanced symmetry in $\mathscr{M}$ then we can use this relationship to describe that background in terms of a Hilbert space of fields at the point of enhanced symmetry.  This seems reminiscent of \cite{Evans:1995su} and indeed we can see that, for geometric backgrounds, this is the same construction as found there. This construction gives a framework in which to describe fibrewise T-duality in torus bundles, the subject of section \ref{s: torus bundle examples}.

\subsection{Twisting the Hamiltonian}

The worldsheet Hamiltonian (density) of a theory related by a geometric duality twist is
\begin{equation}\label{H}
{\cal H}(X)=\left(\begin{array}{cc}
\Pi, & X'
\end{array}\right)\left(\begin{array}{cc}
e^{-1} & 0 \\
-Be^{-1} & e
\end{array}\right){\cal H}_0\left(\begin{array}{cc}
e^{-T} & e^{-T}B \\
0 & e^T
\end{array}\right)\left(\begin{array}{c}
\Pi \\ X'
\end{array}\right),
\end{equation}
where ${\cal H}_0$ is the Hamiltonian density at a point of enhanced symmetry. Here the twist is composed as a product of an $SL(d;\Z)$ transformation and a $B$-field shift, which may be written as
\begin{equation}
\left(\begin{array}{cc}
e^{-1} & 0 \\
-Be^{-1} & e
\end{array}\right)=\left(\begin{array}{cc}
1 & 0 \\
-B & 1
\end{array}\right)\left(\begin{array}{cc}
e^{-1} & 0 \\
0 & e
\end{array}\right).
\end{equation}
For example, for the $SU(2)$ enhanced symmetry we set the radii of the circles to be $\sqrt{\alpha'}$ and the $B$-field to be zero. We may write this as
\begin{equation}
{\cal H}_0=\left(\begin{array}{cc}
G & 0\\
0 & G^{-1}
\end{array}\right).
\end{equation}
Then, ${\cal H}(X)={\cal Z}^TG^{-1}{\cal Z}+{\cal X}^TG{\cal X}$, where
\begin{equation}
{\cal Z}_a=e_a{}^i\big(\Pi_i-B_{ij}X'^j\big),  \qquad  {\cal X}^a=e^a{}_iX'^i.
\end{equation}
We can think of $({\cal Z}_a,{\cal X}^a)$ as twisted versions of $(\Pi_i,X'^i)$. Defining $\partial X_a(E)$ and $\bar{\partial} X_a(E)$ by
\begin{equation}
{\cal Z}_a=(\partial X_a(E)+\bar{\partial} X_a(E)),    \qquad  G_{ab}{\cal X}^b=(-\partial X_a(E)+\bar{\partial} X_a(E)),
\end{equation}
one can show that $\partial \widehat{X}_i=e_i{}^a\partial \widehat{X}_a$ may be written as
\begin{equation}\label{2}
\partial X_i(E)=\frac{1}{2}(\partial X_i+\bar{\partial} X_i)+\frac{1}{2}E_{ij}G^{jk}(\partial X_k-\bar{\partial} X_k).
\end{equation}
This is the same relationship between backgrounds as found by \cite{Evans:1995su}. For nongeometric backgrounds, where we can still write the Hamiltonian as a duality twist of a reference background ${\cal H}={\cal O}{\cal H}_0{\cal O}^T$, ${\cal Z}_a$ and ${\cal X}^a$ may not take the form (\ref{2}). We discuss examples of this type in the following section.

\subsection{Degenerating Fibres}

One might consider an example where the curve $\gamma$ passes through a point in the moduli space where the fibres degenerate. An example of this would be for the $T^2$ bundle where the Deligne-Mumford compactification of the moduli space allows us to include points on the boundary corresponding to a cycle in the $T^2$ fibre degenerating. The transformation of the chiral fields is
\begin{eqnarray}
\left(\begin{array}{c}
     \partial \widehat{X}  \\
     \bar{\partial} \widehat{X}
\end{array}\right)=\frac{1}{2}\left(\begin{array}{cc}
   1+EG^{-1}  & 1-EG^{-1}  \\
  1-E^TG^{-1}   & 1+E^TG^{-1}
\end{array}\right)
\left(\begin{array}{c}
     \partial X  \\
     \bar{\partial} X
\end{array}\right).
\end{eqnarray}
We see that, if the fibres degenerate ($E\rightarrow 0$), then the matrix appearing above has a non-trivial kernel and we can no longer transport the chiral fields past this degeneration (the transformation is no longer invertible). 

A virtue of this framework is that the emphasis is placed upon the integrability of the connection along a path connecting two backgrounds, rather than the existence of globally defined compact isometries. The connection of \cite{Kugo:1992md}, defined on $\mathscr{M}$, is flat and so we can connect all torus backgrounds by parallel transport.

An interesting example where such degenerations are important is the $SU(2)_n$ WZW model\footnote{We thank the referee for bringing this example to our attention.}, which can be formulated as a degenerating torus bundle over an interval with metric \cite{Schulz:2011ye}
\begin{equation}
    ds^2_{S^3} = \frac{1}{4}(d\phi^1)^2+\sin^2\left(\frac{\phi^1}{2}\right)(d\xi^2)^2+\cos^2\left(\frac{\phi^1}{2}\right)(d\xi^3)^2,
\end{equation}
where $\phi^1\in[0,\pi], \xi^{2,3}\in[0,2\pi]$. The full conventions can be found in \cite{Schulz:2011ye}, but we can see that the torus fibres of this metric degenerate at $\phi^1 = 0,\pi$, where one of the circles shrinks to zero. Although such examples would be interesting to understand better, in this paper we shall restrict ourselves to cases where this does not happen.

\section{Torus Bundle Examples} \label{s: torus bundle examples}

We start with some definitions. Consider the $T^d$ bundle over $S^1$ with monodromy $e^N\in O(d,d;\Z)$. $X(\sigma)$ is the base coordinate and $a=1,2...d$ indexes the fibre directions\footnote{Such constructions have received much attention as toy models to study duality and to address issues of moduli stabilization in flux compactifications. See \cite{Kachru:2002sk, Shelton:2005cf, Kaloper:1999yr, Scherk:1979zr,Dabholkar:2002sy,Hull:2005hk} for further details.}. It is useful to write
\begin{equation}\label{N}
N_A{}^B=\left(\begin{array}{cc}
   f_a{}^b  & K_{ab} \\
   Q^{ab}  & -f_b{}^a
\end{array}\right).
\end{equation}
The fields ${\cal A}_A(\sigma)=(e^{-{N X}})_A{}^B\Pi_B(\sigma)$ are defined as ${\cal A}_A(\sigma)=({\cal Z}_a(\sigma),{\cal X}^a(\sigma))$, where
\begin{equation}\label{ZX}
\begin{gathered}
    {\cal Z}_a=(e^{-NX})_a{}^b\Pi_b+(e^{-NX})_{ab}X'^b,\\
    {\cal X}^a=(e^{-NX})^{ab}\Pi_b+(e^{-NX})^a{}_bX'^b,
\end{gathered}
\end{equation}
and $\Pi_A=(\Pi_a,X'^a)$. Note that the ${\cal A}_A(\sigma)$ are well defined as $X(\sigma)$ commutes with the $\Pi_A$ in the fibres.  In the fibres we have (taking $B_{ab}=0$)
    \begin{equation}\label{dX}
    \Pi_a=\partial X_a+\bar{\partial}X_a,    \qquad  G_{ab}X'^b=-\partial X_a+\bar{\partial}X_a.
    \end{equation}
    It will be useful to define the twisted analogues of $\partial X_a$ and $\bar{\partial}X_a$ as
\begin{equation}
{\cal J}_a=\frac{1}{2}\left({\cal Z}_a-G_{ab}{\cal X}^b\right),   \qquad  \bar{\cal J}_a=\frac{1}{2}\left({\cal Z}_a+G_{ab}{\cal X}^b\right).
\end{equation}
Explicitly,
\begin{eqnarray}\label{J}
{\cal J}_a&=&\frac{1}{2}\Big((e^{-NX})_a{}^b-G_{ac}(e^{-NX})^{cb}\Big)\left(\partial X_b+\bar{\partial}X_b\right)\nonumber\\
&&+\frac{1}{2}\Big((e^{-NX})_{ab}-G_{ac}(e^{-NX})^c{}_b\Big)G^{bd}\left(-\partial X_d+\bar{\partial}X_d\right).
\end{eqnarray}
There are similar expressions for $\bar{\cal J}_a$, which are the natural twisted versions of the $\bar{\partial}X_a$. In the given polarization, we write the Hamiltonian density ${\cal H}={\cal H}^{AB}(X)\Pi_A\Pi_B$ in terms of a metric $g_{ab}$ and $B$-field $B_{ab}$ in the fibres,
\begin{equation}\label{M}
{\cal H}^{AB}=(e^{-N^TX})^A{}_CG^{CD}(e^{-NX})_D{}^B=\left(\begin{array}{cc}
   g^{ab}   & g^{ac}B_{cb} \\
B_{ac}g^{cb}  & g_{ab} +B_{ac}g^{cd}B_{db}
\end{array}
\right).
\end{equation}\label{Hamiltonian}
An interesting object to consider is the chiral stress tensor
\begin{equation}\label{T2}
T(\sigma)=G^{ab}{\cal J}_a{\cal J}_b.
\end{equation}
After a lot of tedious but straightforward algebra and using the fact that $e^{-NX}\in O(d,d)$, one can show that $T(\sigma)$ may also be written as
\begin{equation}\label{T}
T(\sigma)=g^{ab}\partial X_a(E)\partial X_b(E),
\end{equation}
where we recall that $\partial X_a(E)$ is given by (\ref{1}), $E_{ab}=g_{ab}+B_{ab}$ is the background tensor and $g_{ab}$ and $B_{ab}$ are \emph{defined} by (\ref{M}). This is true regardless of whether or not the background admits a global geometric description. If the twist matrix $e^{-N X}$ may be written in the same form as the vielbein for ${\cal H}^{AB}$ (as in \ref{H}), then the background will admit a geometric interpretation globally, otherwise it will not in general. We see that $T(\sigma)$ is the parallel transport of the untwisted chiral stress tensor to the background with a duality twist. The examples we consider in this section are toy models but they do play a role in building honest string backgrounds, where there is reason to believe a CFT description exists. In such cases the stress tensor plays an important role in defining the CFT. Here we restrict our attention to toy models and consider\footnote{Similar sentiments hold also for $\bar{T}(\sigma)=G^{ab}\bar{\cal J}_a\bar{\cal J}_b$.} $T(\sigma)$ and, in particular, how it transforms under T-duality given by the automorphism $T(\sigma)\rightarrow e^{iQ}T(\sigma)e^{-iQ}$.

\subsection{The Nilfold}

We study the duality sequence involving the nilfold. This has also been studied in \cite{Hull:2009sg}. The nilfold metric is a $T^2$ bundle over $S^1$ with monodromy $e^{-f}$, where
\begin{equation}\label{f}
f_a{}^b=\left(\begin{array}{cc}
0 & m  \\
0 & 0 
\end{array}\right)
\end{equation}
and the metric is 
\begin{equation}
g_{ij}=\left(\begin{array}{ccc}
1 & 0 & 0 \\
0 & 1 & -mX \\
0 & -mX & 1+(mX)^2
\end{array}\right).
\end{equation}
The $B$-field is taken to be zero. Following (\ref{ZX}), we introduce ${\cal Z}_a=(e^{-f X})_a{}^b\Pi_b$ in the fibres,
\begin{equation}
{\cal Z}_x(\sigma)=\Pi_x(\sigma),  \qquad {\cal Z}_y(\sigma)=\Pi_y(\sigma), \qquad  {\cal Z}_z(\sigma)=\Pi_z(\sigma)+mX(\sigma)\Pi_y(\sigma),
\end{equation}
where $\Pi_i$ are the momenta\footnote{These are also momenta of the untwisted backgrounds - they are conjugate to the coordinates on the fibres. The universality of $(\Pi_i(\sigma),X'^i(\sigma))$ means that we identify the two sets of momenta; $\Pi_i(\sigma)|_E=\Pi_i(\sigma)|_G$.}. Note that $[X(\sigma),\Pi_y(\sigma)]=0$, so these objects are well defined. A rationale for introducing the ${\cal Z}_a$ is the $O(d,d)$ covariant form of the Hamiltonian,
\begin{equation}\label{HamiltonianZX}
{\cal H}(\sigma)={\cal S}_I(\sigma){\cal H}^{IJ}(\sigma){\cal S}_J(\sigma)={\cal A}_I(\sigma)G^{IJ}{\cal A}_J(\sigma),
\end{equation}
with ${\cal A}_I=(e^{N X})_I{}^J{\cal S}_J$, where
${\cal S}_I:=\left(\Pi_i,X'^i\right)$, and the monodromy has been included explicitly and ${\cal A}_I=({\cal Z}_a,{\cal X}^a)$. Here, $G^{IJ}$ denotes (\ref{M}) evaluated at the self-dual background and may be taken to be proportional to the identity. The ${\cal Z}_a$ obey the Heisenberg-like (loop) algebra
\begin{equation}
[{\cal Z}_x(\sigma),{\cal Z}_z(\sigma')]=-2\pi  im\delta(\sigma-\sigma'){\cal Z}_y(\sigma'), \qquad  [{\cal Z}_y(\sigma),{\cal Z}_z(\sigma')]=0,  \qquad  [{\cal Z}_y(\sigma),{\cal Z}_x(\sigma')]=0,
\end{equation}
by virtue of the standard canonical commutation relations on the torus fibres. If we define $\partial X_i$ and $\bar{\partial} X_i$ as in (\ref{dX}), then the ${\cal Z}_a$ of (\ref{ZX}) may be written as
\begin{equation}
{\cal Z}_x(\sigma)=\partial X+\bar{\partial} X,  \qquad
{\cal Z}_y(\sigma)=\partial Y+ \Bar{\partial}Y, \qquad
{\cal Z}_z(\sigma)=\partial Z+\bar{\partial} Z+mX(\partial Y + \Bar{\partial}Y).
\end{equation}
Using the relationship (\ref{1}) and introducing the shorthand $ \partial X_i(E)\equiv\partial \widehat{X}_i$ (i.e. hatted operators are at a general background and unhatted operators are at the $E=G$ background), the change in the fields in going from the background with metric $G_{ij}=\delta_{ij}$ to the nilfold background is
\begin{equation}\label{nilfoldTransformation}
\begin{gathered}
\partial\widehat{X}=\partial X, \qquad  \partial\widehat{Y}=\partial Y-\frac{1}{2}mX(\partial Z-\bar{\partial} Z),\\
\partial\widehat{Z}=\left(1+\frac{1}{2}(mX)^2\right)\partial Z-\frac{1}{2}(mX)^2\bar{\partial} Z-\frac{1}{2}mX(\partial Y-\bar{\partial} Y),
\end{gathered}
\end{equation}
where $X=X_L+X_R$ is unchanged (it is universal).  The stress tensor is given by (\ref{T}) with $\partial X(E)_i\equiv \partial\widehat{X}_i$ given by (\ref{nilfoldTransformation}). Similarly,
\begin{equation}\label{nilfoldAntihol}
\begin{gathered}
\bar{\partial}\widehat{X}=\bar{\partial} X, \qquad  \bar{\partial}\widehat{Y}=\bar{\partial} Y+\frac{1}{2}mX(\partial Z-\bar{\partial} Z),\\
\bar{\partial}\widehat{Z}=\left(1+\frac{1}{2}(mX)^2\right)\bar{\partial} Z-\frac{1}{2}(mX)^2\partial Z+\frac{1}{2}mX(\partial Y-\bar{\partial} Y).
\end{gathered}
\end{equation}
Note that the change in background is a twisting of the torus, i.e. an $SL(2)$ action on the coordinates $(Y,Z)$. Since the ${\cal J}_a$ are $SL(2)$-invariant, they take the same functional form when written using the $\partial X_i$ or the $\partial\widehat{X}_i$. This may be checked explicitly.

\subsubsection{T-Duality}\label{sss: T-duality}

The stress tensor may be written schematically as $T(\sigma)=(\partial \widehat{X})^Tg^{-1}(\partial \widehat{X})$. The T-dual expression is given by $V^{-1}TV$, where $V=e^{-iQ}$, and so, writing $\partial \widehat{X}=U\partial X$,
\begin{equation}
V^{-1}TV=(V^{-1}\partial X^T V)(V^{-1}U^TV)(V^{-1}g^{-1}V)(V^{-1}UV)(V^{-1}\partial X V).
\end{equation}
Assuming $g$ does not contain any dependence on the coordinate we want to dualise along, $(V^{-1}gV)=g$, and then all we need to understand is $V^{-1}UV$. Equivalently, we need to understand $V^{-1}\Gamma V$, i.e. how $\Gamma_I{}^J(X)$ transforms under T-duality. We shall start by studying T-duality along the $Y$ and $Z$ directions. Since $\Gamma_A{}^B=N_A{}^BX(\sigma)$ depends only on $X(\sigma)$, T-duality along the $Y$ and $Z$ directions has no effect and $V^{-1}UV=U$. Thus we need only consider the factor $V^{-1}\partial X V$ which, as discussed at length in section \ref{s: an algebraic approach to T-duality}, is well understood. We see that complications in understanding T-duality arise when $g$ and/or $U$ have explicit dependence on the direction we are performing the duality in. Taken together, the conditions that $V^{-1}UV=U$ and $(V^{-1}g^{-1}V)=g^{-1}$ are that the background is invariant under shifts along the direction in which we are performing the T-duality. The requirement of such invariance is the key ingredient from the Buscher perspective.

The stress tensor for the nilfold is given by $T(\sigma)=g^{ij}\partial \widehat{X}_i\partial \widehat{X}_j$,
\begin{eqnarray}
T(\sigma)&=&(\partial \widehat{X})^2+(\partial \widehat{Y})^2+(\partial \widehat{Z}+mX\partial\widehat{Y})^2\nonumber\\
&=&(\partial X)^2+\left(\partial Y-\frac{1}{2}mX(\partial Z-\bar{\partial}Z)\right)^2+\left(\partial Z+\frac{1}{2}mX(\partial Y+\bar{\partial}Y)\right)^2.
\end{eqnarray}
The T-dual stress tensor is given by $e^{iQ}T(\sigma) e^{-iQ}$, where we perform a T-duality along the $Y$ direction using the charge $Q=\frac{1}{2}\oint d\sigma\cos(2Y_L(\sigma))$. The effect of this automorphism on $Y(\sigma)$ is $e^{iQ}\partial Y(\sigma) e^{-iQ}= -\partial Y(\sigma)$ and $e^{iQ}\bar{\partial} Y(\sigma)e^{-iQ}= \bar{\partial} Y(\sigma)$. The stress tensor of the dual theory is then
\begin{equation}\label{NewT}
T(\sigma)=(\partial X)^2+\left(\partial Y+\frac{1}{2}mX(\partial Z-\bar{\partial}Z)\right)^2+\left(\partial Z-\frac{1}{2}mX(\partial Y-\bar{\partial}Y)\right)^2.
\end{equation}
It is not hard to check that the background is that of a $T^3$ with constant $H$-flux. Using (\ref{dX}) and the expression (\ref{T}),  the stress tensor for the background with $g_{ij}=\delta_{ij}$, $B=mx\rd y\wedge\rd z$ is given by $T(\sigma)=(\partial\widehat{X})^2+(\partial\widehat{Y})^2+(\partial\widehat{Z})^2$, where
\begin{eqnarray}
\partial\widehat{X}=\partial X, \qquad  \partial\widehat{Y}=\partial Y+\frac{1}{2}mX(\partial Z-\bar{\partial}Z),    \qquad  \partial\widehat{Z}=\partial Z-\frac{1}{2}mX(\partial Y-\bar{\partial Y}).
\end{eqnarray}
This is precisely the stress tensor found as the dual of the nilfold stress tensor, as expected. It is more straightforward to construct the related currents ${\cal J}_a$, given by (\ref{J}),
\begin{align}
{\cal J}_x&=\partial X,  \qquad\qquad\qquad\qquad\qquad  \bar{{\cal J}}_x=\bar{\partial} X,\\
{\cal J}_y&=\partial Y-\frac{1}{2}mX(\partial Z-\bar{\partial}Z),   \qquad  \bar{{\cal J}}_y=\bar{\partial}Y+\frac{1}{2}mX(\partial Z-\bar{\partial}Z),\\
{\cal J}_z&=\partial Z+\frac{1}{2}mX(\partial Y+\bar{\partial}Y),    \qquad  \bar{{\cal J}}_z=\bar{\partial} Z+\frac{1}{2}mX(\partial Y+\bar{\partial}Y).
\end{align}
The stress tensor of the nilfold may then be written as (\ref{T2}). 

\subsubsection{A Doubled Algebra}

The twisted versions of the $X'^a$ are given by ${\cal X}^a=(e^{f^TX})^a{}_bX'^b$, where $f^T$ denotes the transpose of (\ref{f}),
\begin{equation}
{\cal X}^x(\sigma)=X'(\sigma), \qquad  {\cal X}^y(\sigma)=Y'-mX(\sigma)Z'(\sigma),    \qquad  {\cal X}^z=Z'(\sigma).
\end{equation}
The ${\cal Z}_a(\sigma)$ and ${\cal X}^a(\sigma)$ close to form an algebra under commutation. The non-trivial commutators are\footnote{Seen by using delta function manipulations and the fact that $(X(\sigma)-X(\sigma'))\delta(\sigma-\sigma')=0$.}
\begin{equation}\notag
[{\cal Z}_x(\sigma),{\cal Z}_z(\sigma')]=-2\pi im\delta(\sigma-\sigma'){\cal Z}_y(\sigma'),    \qquad  [{\cal Z}_x(\sigma),{\cal X}^y(\sigma')]=2\pi im\delta(\sigma-\sigma'){\cal X}^z(\sigma'),
\end{equation}
\begin{equation}
[{\cal Z}_z(\sigma),{\cal X}^y(\sigma')]=-2\pi im\delta(\sigma-\sigma'){\cal X}^x(\sigma'),
\end{equation}
and the central extensions
\begin{equation}
[{\cal Z}_i(\sigma),{\cal X}^j(\sigma')]=2\pi i\delta_i^j\delta'(\sigma-\sigma').
\end{equation}
This algebra is reminiscent of the Lie algebras that appear in flux compactification of supergravity on the nilfold \cite{Hull:2005hk,Hull:2006tp} and also the description of the nilfold in doubled geometry \cite{Hull:2009sg,Hull:2007jy}. We shall comment on this in section \ref{s:relationship to doubled formalism}.

The commutator algebra may be seen to be a centrally extended analogue of the doubled algebra,
\begin{align}
[{\cal Z}_i(\sigma),{\cal Z}_j(\sigma')]&=-2\pi if_{ij}{}^k{\cal Z}_k(\sigma'),\nonumber\\
[{\cal Z}_i(\sigma),{\cal X}^j(\sigma')]&=2\pi i\delta_i^j\delta'(\sigma-\sigma')-2\pi if_{ik}{}^j{\cal X}^k(\sigma')\nonumber,\\
[{\cal X}^i(\sigma),{\cal X}^j(\sigma')]&=0,
\end{align}
where $f_{zx}{}^y=-f_{xz}{}^y=m$. This may be written in an $O(d,d;\Z)$-covariant way as
\begin{equation}\label{alg}
[{\cal A}_I(\sigma),{\cal A}_J(\sigma')]=2\pi iL_{IJ}\delta'(\sigma-\sigma')-2\pi it_{IJ}{}^K{\cal A}_K(\sigma')\delta(\sigma-\sigma'),
\end{equation}
where ${\cal A}_I=({\cal Z}_a,{\cal X}^a)$, $t_{xz}{}^y=-m$ (and zero otherwise) and $L_{IJ}$ is the invariant of $O(d+1,d+1)$.

\subsection{$T^3$ with $H$-flux}

We consider the case where the monodromy matrix is of the form (\ref{N}) with $f^a{}_b=0=Q^{ab}$ and $K_{yz}=m\in\Z$. Note that the ${\cal J}_i=\partial \widehat{X}_i$ for this background are given by (\ref{1}) (with $g_{ab}=\delta_{ab}$ and $B_{yz}=mX$) or read off from the stress tensor (\ref{NewT}) found by dualising the nilfold,
\begin{eqnarray}
{\cal J}_x&=&\partial X,  \qquad \qquad \qquad \qquad \qquad   \bar{\cal J}_x=\bar{\partial}X,    \nonumber\\
{\cal J}_y&=&\partial Y+\frac{1}{2}mX(\partial Z-\bar{\partial}Z),  \qquad   \bar{\cal J}_y=\bar{\partial}Y-\frac{1}{2}mX(\partial Z-\bar{\partial}Z),\nonumber\\
{\cal J}_z&=&\partial Z-\frac{1}{2}mX(\partial Y-\bar{\partial}Y),  \qquad   \bar{\cal J}_z=\bar{\partial}Z-\frac{1}{2}mX(\partial Y-\bar{\partial}Y),
\end{eqnarray}
which gives the components of ${\cal A}_I(\sigma)$ as
\begin{eqnarray}
\begin{array}{ccccc}
{\cal Z}_x=\Pi_x,  &  & {\cal Z}_y=\Pi_y+mXZ', & &  {\cal Z}_z=\Pi_z-mXY',\\
{\cal X}^x=X', & &  {\cal X}^y=Y', & &  {\cal X}^z=Z'.
\end{array}
\end{eqnarray}
The commutation relations are then\footnote{Using 
\begin{eqnarray}
X(\sigma)\frac{\rd}{\rd\sigma'}\delta(\sigma-\sigma')-X(\sigma')\frac{\rd}{\rd\sigma'}\delta(\sigma-\sigma')&=&\frac{\rd}{\rd\sigma'}\Big((X(\sigma)-X(\sigma'))\delta(\sigma-\sigma')\Big)-X'(\sigma')\delta(\sigma-\sigma')\nonumber\\
&=&-X'(\sigma')\delta(\sigma-\sigma'),
\end{eqnarray}
and also the fact that
\begin{equation}
\frac{\rd}{\rd\sigma'}\delta(\sigma-\sigma')=-\frac{\rd}{\rd\sigma}\delta(\sigma-\sigma'),
\end{equation}
which can be easily seen from the Fourier series representation of the periodic delta function.}
\begin{align}
\notag&[{\cal Z}_i(\sigma),{\cal Z}_j(\sigma')]=-2\pi im\varepsilon_{ijk}{\cal X}^k(\sigma')\delta(\sigma-\sigma'),   \qquad  [{\cal Z}_i(\sigma),{\cal X}^j(\sigma')]=2\pi i\delta'(\sigma-\sigma'),    
\\&[{\cal X}^i(\sigma),{\cal X}^j(\sigma')]=0,
\end{align}
which, again, is of the form (\ref{alg}).

\subsection{T-Fold}

Alternatively, T-duality of the nilfold along the $Z$-direction gives a T-fold. It will be instructive to see how the background arises from the currents ${\cal J}_a$. An automorphism with the duality charge amounts to the exchange $\partial Z\rightarrow -\partial Z$,  $\bar{\partial} Z\rightarrow \bar{\partial} Z$. There is no explicit $Z$-dependence, so we need not worry about how $Z$ transforms. The resulting currents are 
\begin{eqnarray}
{\cal J}_x&=&\partial X,  \qquad \qquad \qquad \qquad \qquad   \bar{{\cal J}}_x=\bar{\partial}X,    \nonumber\\
{\cal J}_y&=&\partial Y+\frac{1}{2}mX(\partial Z+\bar{\partial}Z),  \qquad   \bar{{\cal J}}=_y\bar{\partial}Y-\frac{1}{2}mX(\partial Z+\bar{\partial}Z),\nonumber\\
{\cal J}_z&=&\partial Z-\frac{1}{2}mX(\partial Y+\bar{\partial}Y),  \qquad   \bar{{\cal J}}_z=\bar{\partial}Z+\frac{1}{2}mX(\partial Y+\bar{\partial}Y),
\end{eqnarray}
which gives the fields ${\cal Z}_i$ and ${\cal X}^i$ as
\begin{eqnarray}
\begin{array}{ccccc}
{\cal Z}_x=\Pi_x,  &  & {\cal Z}_y=\Pi_y, & &  {\cal Z}_z=\Pi_z,\\
{\cal X}^x=X', & &  {\cal X}^y=Y'-mX\Pi_z, & &  {\cal X}^z=Z'+mX\Pi_y,
\end{array}
\end{eqnarray}
from which we see that $N_A{}^B$ is of the form (\ref{N}) with all entries zero except $Q^{yz}=m$. The stress tensor is given by
\begin{equation}
T=\sum_i{\cal J}_i{\cal J}_i.
\end{equation}
The Hamiltonian may be constructed from ${\cal Z}_i$ and ${\cal X}^i$ as in \eqref{HamiltonianZX}, and the metric and B-field read off for this background,
\begin{equation}
\rd s^2=\rd x^2+\frac{1}{1+(mX)^2}\Big(\rd y^2+\rd z^2\Big), \qquad B=\frac{mX}{1+(mX)^2}\rd y\wedge \rd z.
\end{equation}
The non-trivial commutation relations for the algebra of the ${\cal Z}_i$ and ${\cal X}^i$ are
\begin{equation}\notag
[{\cal Z}_x(\sigma),{\cal X}^z(\sigma')]=-2\pi im{\cal Z}_y(\sigma')\delta(\sigma-\sigma'),   \qquad [{\cal Z}_x(\sigma),{\cal X}^y(\sigma')]=2\pi im{\cal Z}_z(\sigma')\delta(\sigma-\sigma'), 
\end{equation}
\begin{equation}
\qquad  [{\cal X}^y(\sigma),{\cal X}^z(\sigma')]=2\pi im{\cal X}^x(\sigma')\delta(\sigma-\sigma'),
\end{equation}
and the central extension term
\begin{equation}
[{\cal Z}_i(\sigma),{\cal X}^j(\sigma')]=2\pi i\delta_i^j\delta'(\sigma-\sigma').
\end{equation}
This is the central extension of an algebra with structure constant $Q\indices{_x^{yz}}=m$, as we might expect for the T-fold. Again, this algebra is of the general form (\ref{alg}).

\subsection{On Non-isometric T-duality} \label{ss: on non-isometric T-duality}

In this section, we briefly consider a simple case where, in the language of section \ref{sss: T-duality}, $(V^{-1}UV)\neq U$. This can occur when $U$ depends explicitly on the direction we are performing the duality. Duality involving functions of $\partial X$ and $\bar{\partial}X$ are well understood. What if we have a more general function of $X(\sigma)$? The only such functions that appear at the self-dual radius are of the form $e^{inX_L(\sigma)}$, which we have already studied and transform in a well-defined way. However, in considering the obstacles one might need to overcome to apply the operator formalism to non-isometric torus fibrations, it may be instructive to study how functions of $X(\sigma)$ that are not invariant under isometries transform. As a first step we compute $e^{iQ} X_L(\sigma) e^{-iQ}$. In the framework presented here, $X(\sigma)$ is a universal coordinate. Thus, if we know how $X(\sigma)$ transforms under T-duality at the self-dual radius, we can infer how it transforms in backgrounds related to that one by parallel transport. It is not hard to show that
\begin{equation}
    \left[X^i _L(\sigma), X^j _L(\sigma')\right] = i\pi\Theta(\sigma-\sigma')\delta^{i j },
\end{equation}
where\footnote{Note that
\begin{equation}
\frac{\rd}{\rd\sigma}\Theta (\sigma-\sigma')=\frac{1}{2\pi}\sum_ne^{in(\sigma-\sigma')},
\end{equation}
which is the periodic delta-function.}
\begin{equation}
\Theta (\sigma-\sigma')=\frac{1}{2\pi}(\sigma-\sigma')-\frac{i}{2\pi}\sum_{n\neq 0}\frac{1}{n}e^{in(\sigma-\sigma')}.
\end{equation}
The fact that $\Theta$ is not a periodic function and so is not well defined on the worldsheet will be the source of the difficulty in making sense of applying the T-duality automorphism to $X_L(\sigma)$. Using the charge 
\begin{equation}
    Q = \frac{1}{2}\oint \rd\sigma \cos(2X_L(\sigma)),
\end{equation}
with a little work one finds 
\begin{align}
    \notag[Q, X_L(\sigma)] &=  -i\pi \oint \rd\sigma' \Theta(\sigma'-\sigma)\sin(2X_L(\sigma')),\\
    \notag[Q^{(2)}, X_L(\sigma)] &= -\pi^2 \oint \rd\sigma' \Theta(\sigma'-\sigma)X'_L(\sigma'),
\end{align}
where we use the notation 
\begin{equation}\label{nestedCommutators}
    [Q^{(n)},X_L]\equiv [Q,[Q,...[Q,X_L]]...],
\end{equation}
where there are $n$ nested commutators on the RHS. We might think that we can do this integral by parts and get rid of the boundary term in the second expression. Assuming this, you would end up with $X_L\rightarrow-X_L$ as the transformation. However, since $X_L$ and $\Theta$ are not periodic, the boundary term does not vanish. The easiest way to compute the integral is in fact to use the mode expansions and do the integrals directly. Doing this, we have
\begin{align}
    \notag[Q^{(2)}, X_L(\sigma)] 
    = -\pi^2\left(-X_L(\sigma) +\frac{1}{2}\left(X_L(0)+X_L(2\pi)\right)\right).
\end{align}
Since the charge acts in the same way regardless of the value of $\sigma$, we can now just write down the successive commutators:
\begin{align}
    [Q^{(3)},X_L(\sigma)]=& i\pi^3\oint d\sigma' \sin(2X_L(\sigma'))(-\Theta(\sigma'-\sigma)+\frac{1}{2}\Theta(\sigma')+\frac{1}{2}\Theta(\sigma'-2\pi)), \\
    \notag[Q^{(4)}, X_L(\sigma)]=  &\pi^4 \left(X_L(\sigma)-\frac{1}{2}(X_L(0)+X_L(2\pi))\right),\\
    \vdots &
\end{align}
There is a repeating pattern and so the full transformation may be written down 
\begin{align}
    \notag e^{iQ}X_L(\sigma)e^{-iQ} =& \left(-X_L(\sigma)+\frac{1}{2}\left(X_L(0)+X_L(2\pi)\right)\right)\left(\frac{\pi^2}{2!}-\frac{\pi^4}{4!}+...\right)\\
    \notag &+\oint d\sigma'\sin(2X_L(\sigma'))\left(-\Theta(\sigma'-\sigma)+\frac{1}{2}\Theta(\sigma')+\frac{1}{2}\Theta(\sigma'-2\pi)\right)\left(\frac{\pi^3}{3!}-\frac{\pi^5}{5!}+...\right)\\
    \notag&+X_L(\sigma)+\pi\oint d\sigma' \sin(2X_L(\sigma'))\Theta(\sigma'-\sigma)\\
    =&-X_L(\sigma)+X_L(0)+X_L(2\pi)+\frac{\pi}{2}\oint d\sigma' \sin(2X_L(\sigma'))\left(\Theta(\sigma')+\Theta(\sigma'-2\pi)\right).
\end{align}
We can use $\Theta(\sigma'-2\pi)=\Theta(\sigma')-1$ and $\text{sgn}(\sigma')=2\Theta(\sigma')-1$ to slightly simplify the last term, so that
\begin{align}
    \notag e^{iQ}X_L(\sigma)e^{-iQ} =-X_L(\sigma)+X_L(0)+X_L(2\pi)+\frac{\pi}{2}\oint d\sigma' \sin(2X_L(\sigma'))\text{sgn}(\sigma').
\end{align}
Note that, if we use the sine charge instead, we arrive at a different result, namely
\begin{equation}
    \notag e^{iQ}X_L(\sigma)e^{-iQ} =-X_L(\sigma)+X_L(0)+X_L(2\pi)-\frac{\pi}{2}\oint d\sigma' \cos(2X_L(\sigma'))\text{sgn}(\sigma').
\end{equation}
Taking an optimistic view of this rather messy result, we note that it is of the form
\begin{equation}
e^{iQ}X_L(\sigma)e^{-iQ}=-X_L(\sigma)+{\cal C},
\end{equation}
where ${\cal C}$ is a constant operator. This operator depends on the charge one uses to perform the duality and points to the fact that such a shift can be removed by a $U(1)_L\times U(1)_R$ gauge transformation. Put another way, this result is suggestive of the possibility of considering the correct action of the duality on $X_L(\sigma)$ as $X_L(\sigma)\rightarrow -X_L(\sigma)$ (equivalently $X(\sigma)\rightarrow \widetilde{X}(\sigma)$), with the shift by ${\cal C}$ dropping out of all gauge-invariant results. Such a proposal has received support from other quarters (see for example \cite{Gregory:1997te, Harvey:2005ab, Dabholkar:2002sy, Hull:2009sg}). Of course, in those cases the emphasis was on performing T-duality in the absence of isometries in the target space. Here, we see that what is required to be able to neglect ${\cal C}$ is the unbroken $U(1)_L$ symmetry acting on $X_L$. This is a particular linear combination of isometry and $B$-field transformation; the diagonal $U(1)_L\subset U(1)_Z\times U(1)_X$, where the isometry is $U(1)_Z$. An obvious argument against this interpretation is that ${\cal C}$ is an operator and will not commute with other generators in the $SU(2)$ gauge symmetry. As such, it seems a poor candidate for a parameter of a translation symmetry. We will discuss this further elsewhere \cite{Mahmood202?}. The transformations may be written as
\begin{equation}
\delta_ZX=2\alpha,	\qquad	\delta_Z\widetilde{X}=0,	\qquad	\delta_ZX_L=\alpha=\delta_ZX_R
\end{equation}
for the isometry and
\begin{equation}
\delta_XX=0,	\qquad	\delta_Z\widetilde{X}=2\tilde{\alpha},	\qquad	\delta_ZX_L=-\tilde{\alpha}=\delta_ZX_R
\end{equation}
for the $B$-field transformation. We see that, if we choose $\alpha=0$ and $\tilde{\alpha}={\cal C}$, i.e. if we couple the T-duality transformation with a $B$-field transformation, then we may have $X_L(\sigma)$ transforming in the expected way, even if there is no isometry in $X(\sigma)$, provided there is $B$-field symmetry present. This is not a concrete prescription, but it suggests that the formalism considered in this paper may admit more general notions of T-duality if generalised to more interesting backgrounds.

As an application, consider T-duality of the T-fold along the $X$-direction, neglecting the presence of ${\cal C}$. We are therefore, somewhat artificially, elevating the variable $X(\sigma)$ along the base circle from a parameter that characterises the bundle (in the spirit of the construction outlined in section \ref{ss: twisting as a relationship between backgrounds}) to a full quantum field. This pushes us out of the realm of toy models that should in principle be part of a bona fide  CFT and into somewhat uncertain territory. Nonetheless, we shall press on. An automorphism with the duality charge amounts to the exchange $\partial X\rightarrow -\partial X$,  $\bar{\partial} X\rightarrow \bar{\partial} X$. From the above, we shall \emph{assume} that $X\rightarrow\tilde{X}$ (i.e. $X_L\rightarrow-X_L$, where we mean $X_L$ at the enhanced symmetry point). 

The resulting currents are
\begin{eqnarray}
{\cal J}_x&=&\partial X,  \qquad\qquad\qquad\qquad\qquad   \Bar{{\cal J}}_x=\bar{\partial}X,    \nonumber\\
{\cal J}_y&=&\partial Y+\frac{1}{2}m\widetilde{X}(\partial Z+\bar{\partial}Z),  \qquad   \bar{{\cal J}}_y=\bar{\partial}Y-\frac{1}{2}m\widetilde{X}(\partial Z+\bar{\partial}Z),\nonumber\\
{\cal J}_z&=&\partial Z-\frac{1}{2}m\widetilde{X}(\partial Y+\bar{\partial}Y),  \qquad   \bar{{\cal J}_z}=\bar{\partial}Z+\frac{1}{2}m\widetilde{X}(\partial Y+\bar{\partial}Y),
\end{eqnarray}
and so the fields ${\cal Z}_i$ and ${\cal X}^i$ are
\begin{eqnarray}
\begin{array}{ccccc}
{\cal Z}_x=\Pi_x,  &  & {\cal Z}_y=\Pi_y, & &  {\cal Z}_z=\Pi_z,\\
{\cal X}^x=X', & &  {\cal X}^y=Y'-m\widetilde{X}\Pi_z, & &  {\cal X}^z=Z'+m\widetilde{X}\Pi_y.
\end{array}
\end{eqnarray}
These can be obtained by a $T^2$ bundle over the dual circle with ${\cal A}_A(\sigma)=(e^{-N\widetilde{X}})_A{}^B\Pi_B(\sigma)$, where $N_A{}^B$ takes the same form as for the T-fold above. The key point is that $X(\sigma)$ is replaced by $\widetilde{X}(\sigma)$.

The non-trivial commutation relations are then
\begin{align}
    \notag&[{\cal X}^x(\sigma), {\cal X}^z(\sigma')]=-2\pi im\delta(\sigma-\sigma'){\cal Z}_y(\sigma'), \\
    \notag&[{\cal X}^x(\sigma), {\cal X}^y(\sigma')] = 2\pi im\delta(\sigma-\sigma'){\cal Z}_z(\sigma'), \\
    &[{\cal X}^z(\sigma), {\cal X}^y(\sigma')] = -2\pi im\delta(\sigma-\sigma'){\cal Z}_x(\sigma'),
\end{align}
and the central extension
\begin{equation}
    [{\cal Z}_i (\sigma), {\cal X}^j (\sigma')] = 2\pi i\delta\indices{_i^j}\delta'_{\sigma'}(\sigma-\sigma').
\end{equation}
This is reminiscent of the expected R-flux algebra \cite{Shelton:2005cf, Hull:2009sg}. To be clear, the above discussion does not in any way provide a proof that the R-flux background is dual to the T-fold. It does, however, demonstrate that this formalism gives rise to similar algebraic structures seen in the supergravity \cite{Hull:2005hk, Hull:2006tp, Dabholkar:2002sy, Kaloper:1999yr} and doubled geometry \cite{Hull:2009sg} discussions of such backgrounds, \emph{assuming} $X(\sigma)\rightarrow \widetilde{X}(\sigma)$. It also makes clearer the assumptions that are required for such backgrounds to map into each other under T-duality, defined as an automorphism of the operator algebra.

\section{Relationship to the doubled formalism} \label{s:relationship to doubled formalism}

Here we discuss the relationship of the formalisms presented here to the doubled formalism \cite{Hull:2009sg}.

\subsection{The doubled formalisms}

We start with the doubled torus bundle with fibre $T^{2d}$ and a base with coordinate $x\sim x+1$. The coordinates in the fibre are written as $\mathbb{X}^A=(z^a,\tilde{z}_a)$. The doubled torus bundle over a circle
\begin{equation}
T^{2d}\hookrightarrow {\cal T}\rightarrow S^1,
\end{equation}
with monodromy $e^{N}\in O(d,d)$, may be thought of as a $2d+1$ dimensional twisted torus \cite{Hull:2004in, Tseytlin:1990nb, Tseytlin:1990va}. That is, a manifold that is locally a group ${\cal G}_{2d+1}$, but is globally of the form ${\cal T}={\cal G}_{2d+1}/\Gamma_{2d+1}$, where $\Gamma_{2d+1}\subset{\cal G}_{2d+1}$ is a discrete (cocompact) group acting from the left such that ${\cal T}$ is compact. ${\cal T}$ is parallelisable and as such has globally defined, left-invariant one forms
\begin{equation}
  P^x=\rd x,   \qquad  {\cal P}^A=(e^{Nx})^A{}_B\rd\mathbb X^B. 
\end{equation}
A metric on ${\cal T}$ is 
\begin{equation}
\rd s^2=\rd x\otimes \rd x+{\cal M}_{AB}(x)\rd \mathbb{X}^A\otimes\rd\mathbb{X}^B,
\end{equation}
where ${\cal M}_{AB}(x)=(e^{Nx})_A{}^CG_{CD}(e^{N^Tx})^D{}_B$ and $G_{AB}$ is the metric on the untwisted torus fibre. The isometries of ${\cal T}$ are generated by the vector fields
\begin{equation}
Z_x=\frac{\partial}{\partial x},    \qquad  T_A=(e^{-Nx})_A{}^B\frac{\partial}{\partial \mathbb{X}^B},
\end{equation}
and close to give the algebra
\begin{equation}\label{Talgebra}
[T_A,T_B]=0,    \qquad  [Z_x,T_A]=-N_A{}^BT_B.
\end{equation}
If we compactify supergravity on ${\cal T}$, the consistent truncation has gauge algebra which includes 
\begin{equation}\label{Xalgebra}
[T_A,T_B]=N_{AB}X^x,    \qquad  [Z_x,T_A]=-N_A{}^BT_B.
\end{equation}
We see that (\ref{Talgebra}) is a contraction of (\ref{Xalgebra}), where the missing generator is $X^x=\partial/\partial \tilde{x}$ and is associated with $B$-field transformations with gauge parameter lying along the base circle \cite{Hull:2009sg}.

The algebra (\ref{Xalgebra}) may be realised as the isometry algebra of the twisted torus ${\cal G}_{2d+2}/\Gamma_{2d+2}$, where ${\cal G}_{2d+2}$ is a $2d+2$ dimensional group and $\Gamma_{2d+2}$ is a cocompact subgroup of ${\cal G}_{2d+2}$ which acts from the left. The left-invariant one forms on ${\cal G}_{2d+2}/\Gamma_{2d+2}$ are
\begin{equation}
  P^x=\rd x, \qquad    Q_x=\rd\tilde{x}+\frac{1}{2}N_{AB}\mathbb{X}^A\rd\mathbb{X}^B,   \qquad  {\cal P}^A=(e^{Nx})^A{}_B\rd\mathbb X^B,
\end{equation}
and a  natural metric on ${\cal G}_{2d+2}/\Gamma_{2d+2}$ is given by 
\begin{equation}
\rd s^2=\rd x\otimes \rd x+{\cal M}_{AB}(x)\rd \mathbb{X}^A\otimes\rd\mathbb{X}^B+Q_x\otimes Q_x.
\end{equation}
We see that the metric depends on \emph{all} of the coordinates of ${\cal T}$ (the $\mathbb{X}^A$ in addition to the base circle) except $\tilde{x}$. In particular, $Q_x$ may depend on all of the $z^a$ and $\tilde{z}_a$ coordinates. The isometry group of the $2d+2$ dimensional twisted torus is generated by the vector fields dual to these one-forms,
\begin{equation}
Z_x=\frac{\partial}{\partial x},    \qquad   X^x=\frac{\partial}{\partial \tilde{x}},   \qquad  T_A=(e^{-Nx})_A{}^B\left(\frac{\partial}{\partial \mathbb{X}^B}-\frac{1}{2}N_{BC}\mathbb{X}^C\frac{\partial}{\partial \tilde{x}}\right).
\end{equation}
A gauge transformation\footnote{$\mathbb{X}^A\rightarrow (e^{Nx})^A{}_B\mathbb{X}^B$.}
brings the vector fields to the convenient form 
\begin{equation}\label{generators}
Z_x=\frac{\partial}{\partial x}+N^A{}_B\mathbb{X}^B\frac{\partial}{\partial \mathbb{X}^A},    \qquad   X^x=\frac{\partial}{\partial \tilde{x}},   \qquad  T_A=\frac{\partial}{\partial \mathbb{X}^A}-\frac{1}{2}N_{AB}\mathbb{X}^B\frac{\partial}{\partial \tilde{x}}.
\end{equation}
In some sense, the $2d+1$ dimensional twisted torus is less fundamental than the $2d+2$ dimensional construction, as the former arises, from the perspective of the underlying Lie algebra, as a contraction of the latter. However, the route from the physical space to ${\cal T}$ is intuitively very clear - the physical space, in any polarisation, is a torus bundle with a given monodromy and the ${\cal T}$ encodes that monodromy geometrically. By contrast ${\cal G}_{d+2}/\Gamma_{2d+2}$ does not follow in an obvious way from the physical construction, making it difficult to generalise to non-parallelisable cases. Specifically, when considering the original construction - a $T^2$ bundle over $S^1$ with monodromy $e^{N}$ - it is far from obvious \emph{a priori} that the metric on ${\cal  G}_{d+2}/\Gamma_{2d+2}$ will depend explicitly on the $\mathbb{X}^A$, whereas the metric on ${\cal T}$ depends \emph{only} on $x$ and is determined by the monodromy.

\subsection{Universal coordinate formalism}

In the previous section we have seen that the the natural objects that relate the Hamiltonian of a given background to that of the background at an enhanced point involve only the monodromy encoded in the Hamiltonian density
\begin{equation}
{\cal H}_{IJ}(X(\sigma))=(\Pi_x(\sigma))^2+(X'(\sigma))^2+{\cal M}^{AB}\Pi_A(\sigma)\Pi_A(\sigma),
\end{equation}
where we have introduced
\begin{equation}
\Pi_A(\sigma)=\frac{1}{\sqrt{2\pi}}\Big(\Pi_a(\sigma) , X'^a(\sigma)\Big)
\end{equation}
in the torus fibres and ${\cal M}^{AB}(x)$ is the inverse of the metric in the fibres of ${\cal T}$, which is determined by the monodromy and depends only on the base coordinate $x$. The $\Pi_I(\sigma)$ obey the commutation relations
\begin{equation}
[\Pi_A(\sigma),\Pi_B(\sigma')]=iL_{AB}\delta'(\sigma-\sigma'),
\end{equation}
by virtue of the canonical commutation relations. The natural action of $O(d,d)$ on the coordinates and momenta of the $T^d$ fibres suggests the natural $O(d,d)$ covariant objects on the bundle\footnote{All products of operators are assumed normal ordered.}
\begin{equation}
{\cal A}_A(\sigma)=(e^{-NX(\sigma)})_A{}^B\Pi_{B}(\sigma).
\end{equation}
The algebra of these objects is given by the commutator
\begin{eqnarray}
[{\cal A}_A(\sigma),{\cal A}_A(\sigma')]&=&(e^{-NX(\sigma)})_A{}^C(e^{-NX(\sigma')})_B{}^D[\Pi_C(\sigma),\Pi_D(\sigma')]\nonumber\\
&=&i(e^{-NX(\sigma)})_A{}^CL_{CD}(e^{-NX(\sigma')})_B{}^D\delta'(\sigma-\sigma').
\end{eqnarray}
Integrating by parts in $\sigma$ and using the fact that $L_{AB}$ is invariant under the action of $e^{-NX(\sigma)}$ gives
\begin{equation}
[{\cal A}_A(\sigma),{\cal A}_A(\sigma')]=iL_{AB}\delta'(\sigma-\sigma')-iN_{AB}X'(\sigma)\delta(\sigma-\sigma'),
\end{equation}
where $N_{AB}=N_A{}^CL_{CB}=-N_{BA}$. Following the previous section, it is useful to split the fibre fields as ${\cal A}_A(\sigma)=({\cal Z}_a,{\cal X}^a)$ and similarly for the base circle.  The only other non-trivial commutator is $[\Pi_x(\sigma),{\cal A}_A(\sigma')]$, which is easily evaluated to give the algebra
$$
[{\cal A}_A(\sigma),{\cal A}_A(\sigma')]=iL_{AB}\delta'(\sigma-\sigma')-iN_{AB}{\cal X}(\sigma')\delta(\sigma-\sigma'),
$$
$$
[{\cal Z}_x(\sigma),{\cal A}_A(\sigma')]=iN_A{}^B{\cal A}_B(\sigma')\delta(\sigma-\sigma'),    \qquad{}    [{\cal X}^x(\sigma),{\cal A}_A(\sigma')]=0,
$$
\begin{equation}
[{\cal X}^x(\sigma),{\cal Z}_x(\sigma')]=i\delta(\sigma-\sigma').
\end{equation}
This is a central extension of the loop algebra based on the Lie algebra (\ref{Xalgebra}). The interesting fact is that it arises in a very direct way from the intuitive torus bundle geometry. The algebra may be written compactly as
\begin{equation}\label{dubalg}
[{\cal A}_M(\sigma),{\cal A}_N(\sigma')]= iL_{MN}\delta'(\sigma-\sigma')- it_{MN}{}^P{\cal A}_P(\sigma')\delta(\sigma-\sigma'),
\end{equation}
where $t_{xA}{}^B=N_A{}^B$.

\subsection{Zero modes and geometry}

The zero modes of the fibre fields are 
\begin{equation}
X'^a(\sigma)=\omega^a+...,  \qquad{}    \Pi_a(\sigma)=p_a+...,
\end{equation}
where $\omega^a$ is the (dimensionless) winding and $p_a$ is the (dimensionless) momentum zero mode. The ellipsis denote terms with non-trivial $\sigma$-dependence. The coordinates conjugate to these zero modes are $\tilde{z}_a$ and $z^a$ respectively and we can write 
\begin{equation}
X'^a(\sigma)=-i\frac{\partial}{\partial \tilde{z}_a}+...,  \qquad{}    \Pi_a(\sigma)=-i\frac{\partial}{\partial z^a}+...,
\end{equation}
and $\Pi_A(\sigma)=-i\partial_A+...$. Thus, the zero modes of the fields are given by
\begin{equation}
{\cal A}_A(\sigma)=-i(e^{-N X(\sigma)})_A{}^B\frac{\partial}{\partial \mathbb{X}^B}+...,   \qquad{}   {\cal Z}_x(\sigma)=-i\frac{\partial}{\partial x}+..., \qquad{}    {\cal X}^x(\sigma)=-i\frac{\partial}{\partial \tilde{x}}+....
\end{equation}
Truncating to the zero modes (neglecting the $+...$ terms) gives a set of vector fields that generate the isometry algebra (\ref{Talgebra}) with an additional $U(1)$ factor corresponding to the isometry around the dual circle with coordinate $\tilde{x}$. This is a contraction of the algebra (\ref{Xalgebra}). It is interesting to see that it is the non-trivial $\sigma$-dependence that gives rise to the full doubled algebra. Put another way, it is the extended nature of the string that takes us from the algebra (\ref{Talgebra}) that we might expect from particle mechanics on the geometry ${\cal T}\times \widetilde{S}^1$ to the full doubled geometry corresponding to the algebra (\ref{Xalgebra}).

\section{Beyond Torus Bundles} \label{s: beyond torus bundles}

Our focus in this section is on seeing how the framework described in previous sections might generalise to backgrounds that are not necessarily torus bundles. The character of this section will be formal and rather speculative and we do not consider explicit examples, although it would be interesting to do so and there is a clear connection with the flux compactifications of \cite{Kaloper:1999yr,Hull:2006tp}. We do not need to worry about normal ordering issues and a more careful treatment may alter the general scheme outlined here. We shall see that the general algebraic structure that mirrors that of doubled geometry seems to emerge in very general cases. The absence of a detailed understanding of worldsheet theories on such backgrounds makes this section necessarily schematic. In particular, we do not consider normal ordering issues that might arise or potential $\alpha'$ corrections which would alter the form of the Hamiltonian. We do find that algebraic structures reminiscent of (parallelisable) flux compactifications of supergravity \cite{Hull:2006tp} emerge in this approximation; however, unlike those cases, these constructions do not seem to be limited by the requirement that the doubled geometry be locally a group manifold. They are perhaps more reminiscent of compactifications inspired by generalised complex geometry \cite{Hitchin:2010qz, Gualtieri2004Thesis} and its generalisations \cite{Hull:2007zu, Pacheco:2008ps}.

The starting point is the Hamiltonian at the point of enhanced symmetry, denoted by $H(p_o)$,
\begin{equation}
H(p_o)=\oint\rd\sigma\, {\cal S}(p_o){\cal H}(p_o){\cal S}^T(p_o),
\end{equation}
where the point $p_o$ on the space of backgrounds $\mathscr{B}$ is an enhanced point. The generalised metric may be written in terms of generalised vielbeins ${\cal H}(p_o)={\cal E}(p_o){\cal E}^T(p_o)$ as in (\ref{H}) and ${\cal S}_I(p_o;\sigma)=(\Pi_i(\sigma),X'^i(\sigma))_{p_o}$.

The Hamiltonian density at a point $p\in\mathscr{B}$ could be given in terms of the generalised metric
\begin{equation}
{\cal H}(p)={\cal U}(p,p_o){\cal H}(p_o){\cal U}^T(p,p_o),
\end{equation}
where ${\cal U}(p,p_o)={\cal E}(p){\cal E}^{-1}(p_o)$. The Hamiltonian for the theory at $p$ may then be written as
\begin{eqnarray}
H(p)&=&\oint\rd\sigma\, {\cal S}{\cal U}(p,p_o){\cal H}(p_o){\cal U}^T(p,p_o){\cal S}^T\nonumber\\
&=&\oint\rd\sigma\,{\cal A}(p){\cal H}(p_o){\cal A}^T(p).
\end{eqnarray}
Since ${\cal S}_I$ is taken to be a universal coordinate, ${\cal S}_I(p;\sigma)={\cal S}_I(p_o;\sigma)$, and so we can drop the explicit $p$-dependence, and we have defined ${\cal A}(p)={\cal S}(p_o){\cal U}(p,p_o)$.

A polarization is a (maximally isotropic) choice of splitting ${\cal S}$ into $\Pi$ and $X'$. Similarly, we define a polarization of ${\cal A}_I(\sigma)$ as a splitting ${\cal A}_I(\sigma)=\Big({\cal Z}_i(\sigma),{\cal X}^i(\sigma)\Big)$.

\subsection{Flux Compactification on a Twisted Torus} \label{ss: flux compactifications on a twisted torus}

To begin, we consider the example, familiar from many supergravity constructions, of a constant $H$-flux on a parallelisable\footnote{The background need only locally be a group. More generally it could be the quotient of a group by a cocompact subgroup \cite{Hull:2005hk}.} background.  Consider a background that is generated from the reference background by the action of a geometric subgroup of $O(d,d;\Z)$, i.e.
\begin{equation}
\left(
\begin{array}{cc}
e & Be^{-T} \\
0 & e^{-T}
\end{array}
\right),
\end{equation}
where $e\in SL(d)$. If the reference background is the identity, then the metric and $B$-field for this background are
\begin{equation}
g_{ij}=\delta_{ab}e^a{}_ie^b{}_j,	\qquad	B=\frac{1}{2}B_{ij}\rd x^i\wedge \rd x^j.
\end{equation}
The twisted torus with constant flux is a simple example of such a background. Let us take, at the point $p$, the metric and $H$-field to be
\begin{equation}
\rd s^2=\delta_{mn}e^m\otimes e^n,	\qquad	H=\frac{1}{6}K_{mnp}e^m\wedge e^n\wedge e^p,
\end{equation}
where $e^m=e^m{}_i\rd x^i$ is a left-invariant one-form for the group manifold $G$ with structure constants $f_{mn}{}^p=-f\indices{_{nm}^p}$, i.e.
\begin{equation}\label{MC}
\rd e^m+\frac{1}{2}f_{np}{}^me^n\wedge e^p=0.
\end{equation}
The condition $\rd H=0$ then requires $K_{[mn|p}f_{|qt]}{}^p=0$ \cite{Kaloper:1999yr}.

\subsubsection{The Doubled Algebra}

With this polarization, we have
\begin{equation}\label{twistedTorusGenerators}
{\cal Z}_a(\sigma)=(e^{-1})_a{}^i\left(\Pi_i(\sigma)-B_{ij}X'^j(\sigma)\right),	\qquad	{\cal X}^a(\sigma)=e^a{}_iX'^i(\sigma),
\end{equation}
which we may think of as
\begin{equation}
{\cal Z}_a(\sigma)=(e^{-1})_a{}^i\left(-i\frac{\delta}{\delta X^i(\sigma)}-B_{ij}X'^j(\sigma)\right),	\qquad	{\cal X}^a(\sigma)=e^a{}_iX'^i(\sigma).
\end{equation}
Using the canonical commutation relations, we have the algebra (details of the calculations in Appendix \ref{a: details of commutation relations})
\begin{align}
[{\cal Z}_a(\sigma),{\cal Z}_b(\sigma')]&=-f_{ab}{}^c{\cal Z}_c(\sigma')2\pi i\delta(\sigma-\sigma')-K_{abc}{\cal X}^c(\sigma')2\pi i\delta(\sigma-\sigma'),\notag\\
[{\cal Z}_a(\sigma),{\cal X}^b(\sigma')]&=f_{ac}{}^b{\cal X}^c(\sigma')2\pi i\delta(\sigma-\sigma')+\delta_a^b2\pi i\delta'(\sigma-\sigma'),\notag\\
\label{twistedTorusAlgebra}
[{\cal X}^a(\sigma),{\cal X}^b(\sigma')]&=0,
\end{align}
which is of the form (\ref{alg}).

We next verify here that the algebras that we work with are indeed associative. We also discuss generalisations of the above to (non-constant) structure functions, and we show that, at least for geometric flux compactifications, associativity is preserved. 

We compute (details in Appendix \ref{a: details of commutation relations})
\begin{align}
    \notag[{\cal Z}_a(\sigma),[{\cal Z}_b(\sigma'), {\cal Z}_c(\sigma'')]]
    \notag =& -4\pi^2 \delta(\sigma-\sigma'')\delta(\sigma'-\sigma'')\Big(f\indices{_{bc}^d}f\indices{_{ad}^e}{\cal Z}_e(\sigma'')+f\indices{_{bc}^d}K_{ade}{\cal X}^e(\sigma'') \\
    & - f\indices{_{ae}^d}K_{bcd}{\cal X}^e(\sigma'')\Big) + 4\pi^2 \delta'(\sigma-\sigma'')\delta(\sigma'-\sigma'')K_{bcd}\delta_a^d.\label{twistedTorusAssociatorZZZ}
\end{align}
Now we now sum over cyclic permutations. Care must be taken as we are also moving around the $\sigma, \sigma', \sigma''$ dependence. In the first term, this doesn't matter since the delta functions are only supported on $\sigma=\sigma'=\sigma''$, so we can just antisymmetrise on the indices $a,b,c$ without any issues. However, for the second term, since there is a derivative of a delta function and the contributing terms do not appear on the same footing, we have to be more careful. We get
\begin{align}\label{A1}
    \notag &\frac{1}{3}\Big([{\cal Z}_{a}(\sigma),[{\cal Z}_b(\sigma'), {\cal Z}_{c}(\sigma'')]] + \text{cyclic} \Big)\\
    \notag =&  4\pi^2 \delta(\sigma-\sigma'')\delta(\sigma'-\sigma'')\left(f\indices{_{d[a}^e}f\indices{_{bc]}^d}{\cal Z}_e(\sigma'')-2K_{d[ea}f\indices{_{bc]}^d}{\cal X}^e(\sigma'')\right) \\
    &+ \frac{4}{3}\pi^2 K_{abc}\left(\delta'(\sigma-\sigma'')\delta(\sigma'-\sigma'')+\delta'(\sigma'-\sigma)\delta(\sigma''-\sigma)+\delta'(\sigma''-\sigma')\delta(\sigma-\sigma')\right).
\end{align}
The last line here vanishes by delta function manipulations. In order to establish that the other terms vanish, we compute the constraints that arise from $d^2e^m=0$ and the fact that $dH=0$. Taking the exterior derivative of (\ref{MC}) gives $f\indices{_{b[c}^a}f\indices{_{de]}^b}=0$. Similarly, $dH=0$ gives $K_{a[bc}f\indices{_{de]}^a}=0$. Taken together, these two constraints tell us that the right hand side of (\ref{A1}) vanishes.

Since the ${\cal X}^a$ commute with each other, the only other case we have to consider is 
\begin{equation}
    [{\cal X}^a(\sigma),[{\cal Z}_b(\sigma'), {\cal Z}_c(\sigma'')]]+ \text{cyclic}.
\end{equation}
We find that, after a short calculation, 
\begin{align}
    \notag [{\cal Z}_b(\sigma'),[{\cal Z}_c(\sigma''), {\cal X}^a(\sigma)]]
    =&-4\pi^2\delta(\sigma'-\sigma)\delta(\sigma''-\sigma)\left(f\indices{_{cd}^a}f\indices{_{be}^d}{\cal X}^e(\sigma)\right) \\
  &  - 4\pi^2 \delta'(\sigma'-\sigma)\delta(\sigma''-\sigma)f\indices{_{cb}^a}.\label{twistedTorusAssociatorZZX}
\end{align}
When we add the cyclic permutations, the first term will vanish by the Maurer Cartan equation constraint \eqref{MC}. The second term vanishes by delta function manipulations, as with the previous case. Thus, we find that
\begin{equation}
[{\cal X}^a(\sigma),[{\cal Z}_b(\sigma'), {\cal Z}_c(\sigma'')]]+[{\cal Z}_b(\sigma'),[{\cal Z}_c(\sigma''), {\cal X}^a(\sigma)]]+[{\cal Z}_c(\sigma''),[{\cal X}^a(\sigma), {\cal Z}_b(\sigma')]]=0,
\end{equation}
and so the algebra is indeed associative.

\subsection{From Structure Constants to Structure Functions} \label{ss: from structure constants to structure functions}

We would like to see how far we can generalise this, and in particular we would like to see if we can relax the condition that $f_{ab}{}^c$ and $K_{abc}$ are constant and allow them to be functions
\begin{equation}
f_{ab}{}^c\rightarrow f_{ab}{}^c(X), 	\qquad	K_{abc}\rightarrow K_{abc}(X).
\end{equation}
In particular, we take $g_{ij}$ and $B_{ij}$ to be general (we assume the metric is torsion-free). It is fairly easy to see that the algebra will still go through without any problems. This is essentially because there are no derivatives of $f$ or $K$ in the derivation of the algebra. However, where there might be problems is associativity. Since associators have nested commutators, we \textit{do} have derivatives of $f$ and $K$ appearing. However, we will find that the algebra is in fact still associative. This is in contrast to the doubled geometry construction, where the group structure plays a prominent role. However, we suspect a formal generalisation of doubled geometry along similar lines is possible. Imposing the self-duality constraint there might require gauging an algebroid structure along the lines of \cite{Kotov:2010wr} and we shall not comment on this further here.

Firstly, we should derive the modified constraints that now arise from the Maurer-Cartan equation (\ref{MC}) and the flux condition $\rd H=0$. Going through the same process we find that 
\begin{equation}\label{MCConstraintNon-constant}
e\indices{_{[a}^i}\partial_i f\indices{_{bc]}^d} - f\indices{_{e[ ab}^d}f\indices{_{bc]}^e}=0, \qquad	e\indices{_{[a}^i}\partial_i K_{bcd]} - \frac{3}{2}K_{e[ ab}f\indices{_{cd]}^e}=0.
\end{equation}
How does this alter the calculations checking associativity? Essentially, there are two changes to consider: the explicit derivatives of $f$ and $K$, and the changes to the $\delta'$ terms (which are themselves a result of the $\sigma$ dependence of $f$ and $K$). For example, we now have 
\begin{align}
    \notag [{\cal Z}_a(\sigma),[{\cal Z}_b(\sigma'), {\cal Z}_c(\sigma'')]]
    =& -4\pi^2 \delta(\sigma-\sigma'')\delta(\sigma'-\sigma'')\left\{f\indices{_{bc}^d}f\indices{_{ad}^e}{\cal Z}_e(\sigma'')+f\indices{_{bc}^d}K_{ade}{\cal X}^e(\sigma'')\right. \\
    \notag &\left. - f\indices{_{ae}^d}K_{bcd}{\cal X}^e(\sigma'')+ e\indices{_a^i}\partial_i f\indices{_{bc}^d}{\cal Z}_d(\sigma'')+e\indices{_a^i}\partial_iK_{bce}{\cal X}^e(\sigma'')\right\} \\
    \label{ZZZassociatorSingle}&+ 4\pi^2 \delta'(\sigma-\sigma'')\delta(\sigma'-\sigma'')K_{bcd}(\sigma'')\delta_a^d.
\end{align}
The new terms are the last two terms in the braces. Now, when we antisymmetrise this, the term we have to be extra careful with is the last one, i.e. the term outside of the braces. This is because the $K$ now has $\sigma''$ dependence. We can write this term as 
\begin{align}
    \notag -\delta'(\sigma-\sigma'')\delta(\sigma'-\sigma'')K_{bcd}(\sigma'')\delta_a^d  =& \partial_{\sigma''}\left(\delta(\sigma-\sigma'')K_{abc}(\sigma'')\right)\delta(\sigma'-\sigma'') \\ \notag 
    &- \delta(\sigma-\sigma'')\delta(\sigma'-\sigma'')K'_{abc}(\sigma'')\\
    \notag =& -\delta'(\sigma-\sigma'')\delta(\sigma'-\sigma'')K_{abc}(\sigma) \\ 
    &- \delta(\sigma-\sigma'')\delta(\sigma'-\sigma'')X'^i\partial_i K_{abc}(\sigma'').
\end{align}
Thus, after antisymmetrising, the total contribution from the $\delta'$ terms is
\begin{align}
    \notag &\frac{4}{3}\pi^2K_{abc}\left(\delta'(\sigma-\sigma'')\delta(\sigma'-\sigma'')+\delta'(\sigma'-\sigma)\delta(\sigma''-\sigma)+\delta'(\sigma''-\sigma')\delta(\sigma-\sigma')\right) \\
    \notag &+4\pi^2 \delta(\sigma-\sigma'')\delta(\sigma'-\sigma'') \frac{1}{6}e\indices{_e^i}\partial_iK_{abc}{\cal X}^e(\sigma'')\\
    &=4\pi^2 \delta(\sigma-\sigma'')\delta(\sigma'-\sigma'') \frac{1}{3}e\indices{_e^i}\partial_iK_{abc}{\cal X}^e(\sigma'') ,
\end{align}
the first term vanishing as in the constant case. We will see that the remaining term contributes in such a way as to ensure associativity. The full expression can now be written as 
\begin{align}
    \notag [{\cal Z}_{[a}(\sigma),[{\cal Z}_b(\sigma'), {\cal Z}_{c]}(\sigma'')]]
    =& 4\pi^2 \delta(\sigma-\sigma'')\delta(\sigma'-\sigma'')\left\{f\indices{_{e[a}^d}f\indices{_{bc]}^e}{\cal Z}_d(\sigma'')-2K_{d[ea}f\indices{_{bc]}^d}{\cal X}^e(\sigma'')\right. \\
    &\left. - e\indices{_{[a}^i}\partial_i f\indices{_{bc]}^d}{\cal Z}_d(\sigma'')-e\indices{_{[a}^i}\partial_iK_{bc]e}{\cal X}^e(\sigma'')+\frac{1}{3}e\indices{_e^i}\partial_iK_{abc}{\cal X}^e(\sigma'')\right\},
    \label{ZZZassociator}
\end{align}
and we can write the last two terms in this expression as 
\begin{align}
    &-e\indices{_{[a}^i}\partial_iK_{bc]e}{\cal X}^e(\sigma'')+\frac{1}{3}e\indices{_e^i}\partial_iK_{abc}{\cal X}^e(\sigma'') = \frac{4}{3}e\indices{_{[e}^i}\partial_i K_{abc]}{\cal X}^e.
\end{align}
Thus, \eqref{ZZZassociator} becomes
\begin{align}
    \notag [{\cal Z}_{[a}(\sigma),[{\cal Z}_b(\sigma'), {\cal Z}_{c]}(\sigma'')]]
    =& 4\pi^2 \delta(\sigma-\sigma'')\delta(\sigma'-\sigma'')\Big\{f\indices{_{e[a}^d}f\indices{_{bc]}^e}{\cal Z}_d(\sigma'')-2K_{d[ea}f\indices{_{bc]}^d}{\cal X}^e(\sigma'') \\
    & - e\indices{_{[a}^i}\partial_i f\indices{_{bc]}^d}{\cal Z}_d(\sigma'')+\frac{4}{3}e\indices{_{[e}^i}\partial_i K_{abc]}{\cal X}^e(\sigma'')\Big\}=0,
    \label{ZZZassociatorFinal}
\end{align}
where we notice that the expressions in the braces are precisely the constraints imposed by the equations \eqref{MCConstraintNon-constant} and therefore vanish. The calculation of $[{\cal Z},[{\cal Z},{\cal X}]]$ also works out in a similar, though slightly simpler, way. Thus, we conclude that the algebra is still associative even if $f_{ab}{}^c$ and $K_{abc}$ are not constant.

\subsection{A comment on associativity of the R-flux background}

In this section we have focused only on geometric flux compactifications because this is the simplest case to approach in the general setting which we have laid out. It would be interesting to extended to include other, possibly non-geometric, backgrounds but we have not considered this here. For the specific case of the R-flux background discussed earlier, it is straightforward to study the associativity of the algebra and we find that the algebra is indeed associative. The details are similar to those given in appendix \ref{a: details of commutation relations}. There has been much discussion of the R-flux background giving rise to a non-associative structure\footnote{See, for example, \cite{Cornalba:2001sm,Herbst:2001ai,Bouwknegt:2004ap,Blumenhagen:2010hj,Lust:2019hmr}.} but we find no sign of any such structure in our construction.

\section{Discussion} \label{s: discussion}

Our aim in this article was to provide a framework in which to discuss T-duality in a wide range of cases that did not rely solely on questions of the existence of isometries of the target space theory. One motivation for a different approach is the desire to better understand under what conditions a T-dual description of a background exists, given that there are cases where global isometries are not a feature of the background yet something akin to T-duality appears possible. Another motivation was to reframe and place in a contemporary setting the operator algebra arguments around stringy symmetries which appeared in the older literature \cite{Evans:1989xq,Evans:1995su}. That this approach provides a language to discuss T-duality that is not reliant on target space concepts is particularly appealing. To this end we clarified some of the issues surrounding the operator approach to T-duality that were not, to our knowledge, addressed in the older literature. In particular, the non-uniqueness of the T-duality charge and the role isometries played in simplifying the discussion were clarified.

We sketched out a general framework in which issues of T-duality rest on the construction of a connection and a path $\gamma\subset{\mathscr{B}}$ between a background of enhanced symmetry, in which the duality is manifest, and the background in question. This combines the ideas of \cite{Dine:1989vu,Evans:1995su} with the studies of connections on the space of string backgrounds given in \cite{Kugo:1992md,Ranganathan:1992nb,Ranganathan:1993vj}. For on-shell considerations, the connection is on the state space of CFTs, but this can be generalised to more general sigma models. From a string field theory perspective, this provides a way of discussing off-shell physics. This provides a different starting point for T-duality than the traditional Buscher construction and one that may admit concrete discussions of non-isometric generalisation from the perspective of the full worldsheet quantum theory. A concrete universal proposal would rest on the availability of a definition of what the space $\mathscr{B}$, that the connection defines a parallel transport over, is. Limiting ourselves to the space of CFTs clarifies the problem somewhat, but possibly at the expense of ruling out much of what makes the Buscher procedure so useful - the fact that one can apply it to general sigma models without limiting to exact string backgrounds, few of which are known explicitly.

Concrete (off-shell) torus bundle examples were studied. The question of whether the theories are genuinely T-dual rests on the issue of whether the charge is preserved by the Hamiltonian. Non-geometric backgrounds were included. The T-fold considered emerged naturally in this formalism, whereas the `R-flux' background seemed problematic. Specifically, the issue of how $X(\sigma)$ transforms was discussed and it was found that the simple $X(\sigma)\rightarrow \widetilde{X}(\sigma)$ transformation may not be too naive. It would be good to study this issue in the context of an exact solution (or to leading order in $\alpha'$), as is discussed in \cite{Tong:2002rq,Harvey:2005ab,Jensen:2011jna} or more recently the class of backgrounds found in \cite{Chaemjumrus:2019ipx,Chaemjumrus:2019wrt}. As a simpler and more tractable case, it would be interesting to study the duality on orbifolds where some part of the enhanced gauge symmetry is broken by the orbifold action. Again, this might give more perspective on Mirror symmetry in K3 and Calabi-Yau manifolds via their orbifold limits. For similar reasons, it would be interesting to see to what extent the construction considered here could be generalised to torus bundles with degenerating fibres.
    
Somewhat surprisingly, the doubled algebra was shown to arise from the zero modes of the commutation relations of the torus bundle directly, where the central extensions played an important role. The doubled geometry arose in this context as an effective classical description of the quantum theory. The distinction between the doubled geometry of \cite{Hull:2009sg} and the centrally extended algebras that appeared here is subtle and deserves further investigation. Similarly, it would be interesting to know what significance the algebras discussed in section \ref{s: beyond torus bundles} have. They are clearly related to the parallelizable flux compactifications \cite{Kaloper:1999yr,Hull:2006tp}, but may have wider applicability. In the same way that the contraction of the doubled algebra generated by the vector fields (\ref{generators}) arose from the zero modes of the operator algebra (\ref{dubalg}), it would be interesting to see if there are non-parallelisable cases where the operator algebra of section \ref{ss: from structure constants to structure functions} can be used to find a concrete proposal for the corresponding doubled geometry.

Another interesting direction is further investigation of non-abelian T-duality \cite{delaOssa:1992vci,Giveon:1993ai}. It would be interesting to see whether, using the formalism in this paper, the status of non-abelian T-duality could be further clarified. The perspective that the enhanced symmetry group should have some off-shell significance and is broken by a choice of vacuum may be useful here. In the cases we have considered, torus bundles -  for which the action of the unbroken $\Z_2$ symmetry is clear - play a central role. To make progress on the general question of non-abelian duality from the perspective advocated here, one would need a better understanding of the relationship, if any, between the enhanced symmetry group and the non-abelian isometries of the target space.

\begin{center}
\textbf{Acknowledgments}
\end{center}
This work has been partially supported by STFC consolidated grants ST/P000681/1, ST/T000694/1. RR is in part supported by the generosity of the Avery Foundation.

\appendix

\section{Vertex Operators and WZW} \label{a: vertex operators and WZW}

The following sections contain some comments and observations that may be known but, to our knowledge, have not appeared in the literature.

\subsection{More general operators}

In this section we extend the results of previous sections to more general operators with a particular focus on those operators which may be thought of as the building blocks of vertex operators. Of particular interest are the vertex operators, for which we shall need a better understanding of how operators $e^{ik\cdot X}$ and $\partial^nX^{\mu}$ transform under the particular automorphism in question for general $k_{\mu}$ and $n$. In general, the action of an $SU(2)$ automorphism will transform an operator of conformal weight $h$ into a linear combination of other operators of the same weight.  In particular, though the transformation of $\partial X_L(z)$ is straightforward, the transformation of $X_L(z)$ is anything but (see section \ref{ss: on non-isometric T-duality}), and so the transformation of operators of the form $e^{in X_L(z)}$ needs careful consideration. We will use OPEs instead of commutation relations here as they are easier to work with for exponential calculations. Therefore, we will need to use Euclidean signature. 

The transformation of $\partial^nX$ was given in \eqref{higherDerivativeTransformation} and was straightforward to deduce. As mentioned earlier, \cite{Evans:1995su} compute the transformation of exponentials using a point-splitting argument, but we will take an inductive approach.

The transformation of $e^{2iX_L(z)}$ under the T-duality automorphism was considered in section \ref{s: an algebraic approach to T-duality}. In order to better understand the transformation of $e^{inX_L(z)}$ for $n\in\Z$, let us next look at the transformation of $e^{iX_L(z)}$. Using the OPE \eqref{XXOPE}, we have 
\begin{equation}
    [Q, e^{iX_L(w)}] =\frac{\pi}{2}e^{-iX_L(w)}, 
\qquad
    [Q^{(2)}, e^{iX_L(w)}] =\frac{\pi^2}{4}e^{iX_L(w)},
\end{equation}
where the notation for nested commutators is given in \eqref{nestedCommutators}. There is a clear repeating pattern, oscillating between $e^{iX}$ and $e^{-iX}$, and it is not hard to guess the general term. Thus, we obtain the result
\begin{equation}
    e^{iQ} e^{iX_L(z)}e^{-iQ} = i e^{-iX_L(z)}.
\end{equation}
This transformation seems at odds with the general expectation $X_L(z)\rightarrow -X_L(z)$ that we have seen in the massless vertex operators, suggesting instead $X_L(z)\rightarrow -X_L(z)+\pi/2$ (though this isn't true either). In fact, as we will now show, we have
\begin{equation}\label{expGeneraln}
    e^{inX_L(z)} \rightarrow \left\{
    \begin{matrix}
        ie^{-inX_L(z)}, \qquad \text{n odd} \\
        e^{-inX_L(z)}, \qquad \text{n even}.
    \end{matrix}\right.
\end{equation}
We see that, in the $n$ odd case, there is an extra factor of $i$ compared to expectations. It is in fact not the case that we can simply look at transformations such as \eqref{expGeneraln} and deduce a transformation of $X_L$. We can prove \eqref{expGeneraln} via an inductive argument, which goes as follows. Define composite operators $A^{\pm}_n$ by 
\begin{equation}
    e^{\pm i2X_L(z)} = e^{\pm i2X_L(w)}\sum_{n=0}^{\infty} \frac{1}{n!} A_n^{\pm}(w).
\end{equation}
Then, the first step is to show that 
\begin{equation}\label{expCoefficientSplit}
    A_n^{\pm} = E_n\pm O_n,
\end{equation}
where $E_n$ is even under T-duality and $O_n$ is odd. We can prove this via induction. We have:
\begin{align}
    \notag A_n^+ =& 2i\partial X A_{n-1}^+ + \partial A_{n-1}^+\\
    \notag =& 2i\partial X (E_{n-1}+O_{n-1}) + \partial E_{n-1}+\partial O_{n-1} \\
    =&: E_n + O_n, 
\end{align}
where 
\begin{align}
    E_n = 2i\partial X O_{n-1} + \partial E_{n-1}, \\
    O_n = 2i\partial X E_{n-1} + \partial O_{n-1}.
\end{align}
Doing the same for $A_n^-$, we find that 
\begin{equation}
    A_n^- = E_n - O_n,
\end{equation}
as required. Finally, we note that the result is clearly true for $n=1$, and hence we have proven \eqref{expCoefficientSplit}.\\
The next step is to consider the following commutator: 
\begin{align}
    \notag [Q, \cos(nX_L)] =& \frac{1}{8}\oint dz (e^{i2X_L(z)}+e^{-i2X_L(z)})(e^{inX_L(w)}+e^{-inX_L(w)})\\
    \notag \sim& \frac{\pi}{4}\oint dz (z-w)^{-n}\left(e^{-i(n-2)X(w)}\sum_{m}\frac{1}{m!}A_m^+(w)+e^{(n-2)X(w)}\sum_{m}\frac{1}{m!}A_m^-(w)\right)\\
    \notag =& \frac{\pi}{4}\left(e^{-i(n-2)X_L(w)}\frac{1}{(n-1)!}A_{n-1}^+(w)+e^{i(n-2)X_L(w)}\frac{1}{(n-1)!}A_{n-1}^-(w)\right)\\
    =& \frac{1}{2(n-1)!}\left(E_n(w)\cos((n-2)X_L(w))-iO_n(w)\sin((n-2)X_L(w))\right). \label{generalnCommutator}
\end{align}
Now, what we are leading up to is the result
\begin{align}
    \cos(nX_L) \rightarrow \left\{
    \begin{matrix}
        i\cos(nX_L), \qquad \text{n odd}\\
        \cos(nX_L), \qquad \text{n even},
    \end{matrix} 
    \right. \label{generalnCos}\\
    \sin(nX_L) \rightarrow \left\{
    \begin{matrix}
        -i\sin(nX_L), \qquad \text{n odd}\\
        -\sin(nX_L), \qquad \text{n even},
    \end{matrix} 
    \right. \label{generalnSin}
\end{align}
which is equivalent to \eqref{expGeneraln}. Once again, we will use induction, this time via \eqref{generalnCommutator}. First, note that 
\begin{equation}
    [Q, e^{iQ} P e^{-iQ}] = e^{iQ}[Q,P]e^{-iQ},
\end{equation}
where $P$ is any operator. Using this in \eqref{generalnCommutator} with the known transformations of $E_n,O_n$ and the induction hypothesis on $\cos((n-2)X_L), \sin((n-2)X_L)$ (i.e. the RHS of \eqref{generalnCommutator}), we deduce that \eqref{generalnCos} is indeed true (we've already shown that it is true for $n=1,2$). The same process leads to the proof of \eqref{generalnSin}, and hence we have shown that \eqref{expGeneraln} is true, as required.\\
Note that we could also do exactly the same with the sine charge and it would just give different phases for the exponential transformations. We will just state the results: 
\begin{equation}
    e^{niX_L} \xrightarrow{Q^s} \left\{
    \begin{matrix}
        (-1)^{\frac{n+1}{2}}e^{-niX_L}, \qquad \text{n odd}, \\
        (-1)^{\frac{n}{2}}e^{-niX_L}, \qquad \text{n even}.
    \end{matrix}\right.
\end{equation}
This leads to 
\begin{align}
    \cos(nX_L) \xrightarrow{Q^s} \left\{
    \begin{matrix}
        (-1)^{\frac{n+1}{2}}i\sin(nX_L), \qquad \text{n odd},\\
        (-1)^{\frac{n}{2}}\cos(nX_L), \qquad \text{n even},
    \end{matrix} 
    \right. \label{generalnCosSinCharge}\\
    \sin(nX_L) \xrightarrow{Q^s} \left\{
    \begin{matrix}
        (-1)^{\frac{n+1}{2}}i\cos(nX_L), \qquad \text{n odd},\\
        (-1)^{\frac{n}{2}+1}\sin(nX_L), \qquad \text{n even}.
    \end{matrix} 
    \right. \label{generalnSinSinCharge}
\end{align}
We can see that, as opposed to the cosine charge, we have sines and cosines transforming into each other. However, as we saw in \ref{ss: gauge equivalence of T-duality charges}, it doesn't matter which charge we use since they are equivalent up to $U(1)$ gauge transformations. 

\subsection{The WZW formulation}

To better understand what is going on, we turn to the WZW formulation of the model at the self-dual radius.

We can formulate the T-duality of circle compactifications at the SDR using the $SU(2)$ WZW model. The massless $(1,0)$ currents define a level 1 $\widehat{su(2)}$ affine lie algebra. If we define
\begin{equation}
J^1 = \cos(2X_L), \qquad J^2 = \sin(2X_L), \qquad J^3 = i\partial X, 
\end{equation}
then, as with all conformal primaries, we can expand them in modes as 
\begin{equation}
    J^i(z) = \sum_{n=-\infty}^{\infty} J^i_n z^{-n-1}.
\end{equation}
These obey the level 1 current algebra
\begin{equation}
    [J_n^i, J_m^j] = 2\pi\left(\frac{n}{2}\delta_{n+m}\delta^{ij}+i\epsilon^{ijk}J_{n+m}^k\right).
\end{equation}
In this formulation, the (cosine) charge takes the simple form
\begin{equation}
    Q = \frac{1}{2}\oint\rd z\,\cos(2X_L(z)) =  \frac{1}{2}J_0^1,
\end{equation}
or $J^2_0$ for the sine charge. It can then be verified that the effect of the charge on the modes $J^i(z)\rightarrow e^{iQ}J^i(z)e^{-iQ}$ is 
\begin{align}
    J_n^1\rightarrow J_n^1, \qquad
    J_n^2\rightarrow-J_n^2, \qquad
    J_n^3\rightarrow-J_n^3. \label{T-dualityWZWModeTransformation}
\end{align}
This should reproduce all of the T-duality transformation results that we derived earlier. We will verify this using the states corresponding to the relevant vertex operators. The states at level 1 can be constructed from the highest weight states with level 1 by acting on them with the modes $J^i_n$, $n<0$ (note that the level of a weight in the case of $SU(2)$ is given by the sum of its Dynkin labels). The states that are generated from a single highest weight state $|\lambda\rangle$ form the irreducible module $L_\lambda$. In the case of $SU(2)$, it turns out that there are 2 such modules, $L_{[1,0]}$ and $L_{[0,1]}$, where the subscripts are the Dynkin labels of the corresponding highest weight states (see \cite{Franceso:1997} for a review). We start with the module $L_{[1,0]}$, for which the highest weight state is simply the vacuum. The generic state is then
\begin{equation}
    |\lambda'\rangle = J^i_{-n} ... J^j_{-m} |0\rangle.
\end{equation}
For the $L_{[0,1]}$ module, the highest weight state is $|0'\rangle:=e^{iX_L(0)}|0\rangle$, and the generic state is 
\begin{equation}
    |\lambda'\rangle = J^i_{-n} ... J^j_{-m} |0'\rangle.
\end{equation}
Note that, from the point of view of the WZW model, it is clear to see that each mass level has either "odd" or "even" exponentials, but not both (i.e. $e^{inX_L}$ where $n$ is odd or even). The vertex operators with odd and even exponentials belong to different modules, so they don't mix under the action of $SU(2)$. It is only via the WZW formulation that this becomes clear. In particular, the action of T-duality on one of these states is 
\begin{equation}
    e^{iQ}J_{-n}^i...J_{-m}^j|\lambda\rangle = e^{iQ}J^i_{-n}e^{-iQ}...e^{iQ}J^j_{-m}e^{-iQ}e^{iQ}|\lambda\rangle,
\end{equation}
where $|\lambda\rangle$ is one of the highest weight states. If we recall also that, under T-duality with the cosine charge, 
\begin{equation}
    e^{iX_L}\rightarrow ie^{-iX_L},
\end{equation}
we see that the action of T-duality simply amounts to sign changes and the possible factor of $i$ coming from the transformation of the $|0'\rangle$ state,
\begin{equation}
e^{-iQ}|0'\rangle=i|0'\rangle.
\end{equation}
There is no mixing between the two modules\footnote{We can also write 
\begin{equation}
    e^{iX_L}|0\rangle \rightarrow ie^{-iX_L}|0\rangle = iJ^-_{0}e^{iX_L}|0\rangle,
\end{equation}
where $J^{\pm} = J^1\pm i J^2$, and we can expand $J^-$ to get 
\begin{equation}
    iJ^-(z)e^{iX_L}|0\rangle = i\sum_{n} :\frac{J^-_n}{z^{n+1}}e^{iX_L}:|0\rangle \xrightarrow{z\rightarrow 0} iJ^-_{-1}e^{ix_L}|0\rangle,
\end{equation}
where $x_L = \frac{1}{2}(x-\tilde{x})$, and we have used the fact that the normal ordering means we don't have to worry about terms in $J^-$ which don't commute with $e^{iX_L}$. }. This can also be seen from the fact that the states at each grade (i.e. each conformal dimension) form representations of $SU(2)$.

\subsection{An example}

Let's now look at a specific example to see explicitly the equivalence between these two formulations of the bosonic string at the SDR. We will look at the case of the operators
\begin{equation}
    \partial ^2 X(z), \qquad \partial X(z)\partial X(z), \qquad \partial X(z) e^{\pm2iX_L(z)},
\end{equation}
which are used to build massive vertex operators. We can order these by their eigenvalues of $J^3_0=\oint J^3$. The state with the largest eigenvalue is the highest weight state, and we can obtain all other states by acting with $J^-_0=\oint J^-$. For example, we have 
\begin{align}
 [J^3_0,\partial X(w) e^{2iX_L(w)}]=\partial X(w) e^{2iX_L(w)},
\end{align}
i.e. the $J^3_0$ eigenvalue is $+1$ in this case. Repeating for the other states, we find that the eigenvalues for $\partial^2X, \partial X^2, \partial X e^{-2iX_L}$ are $0,0,-1$ respectively. Thus, $\partial X e^{2iX_L}$ corresponds to the highest weight state. Acting with $J^-_0$, we obtain
\begin{align}
  [J^-_0, \partial X(w)e^{2iX_L(w)}] = \oint dz J^{-}(z) \partial X(w)e^{2iX_L(w)} =\partial^2X(w).
\end{align}
Similarly, lowering again gives $\partial X e^{-2iX_L}$. Thus, these three states are in a triplet of $SU(2)$ and $\partial X^2$ is in a singlet. We can also determine the action of T-duality straightforwardly using the state-operator correspondence. The states corresponding to the above operators are \begin{align}
        \partial X(z) e^{\pm 2iX_L(z)} &\leftrightarrow J^3_{-1}J^{\pm}_{-1} |0\rangle \equiv |2,\pm1\rangle_{(2)}, \\
    \partial^2 X(z) &\leftrightarrow J^-_{-2}|0\rangle \equiv |2, 0\rangle_{(2)}, \\
    \partial X^2(z) &\leftrightarrow J^3_{-1}J^3_{-1}|0\rangle \equiv |2, 1\rangle_{(0)}, 
\end{align}
where the numerical labels are, respectively, the $L_0$ eigenvalue, the $J^3_0$ eigenvalue and the $SU(2)$ representation, and normal ordering is implicit. By direct calculation or by using the action of T-duality on the modes given in \eqref{T-dualityWZWModeTransformation}, we see that the states transform in exactly the same way as the operators would imply, i.e. we recover the transformation 
\begin{align}
\partial ^2X \rightarrow -\partial ^2X, \qquad	\partial X^2\rightarrow \partial X^2, \qquad	\partial X e^{\pm2iX_L} \rightarrow -\partial X e^{\mp2iX_L}.
\end{align}

\section{Elliptic Monodromy} \label{a: elliptic monodromy}

We present an example of a simple torus bundle with a geometric twist over the base circle from the elliptic conjugacy class of $SL(2;\Z)$. Specifically, we will look at the monodromy 
\begin{equation}
    f = \begin{pmatrix}
    0&\pi/2 \\ -\pi/2 & 0 
    \end{pmatrix}
    \implies e^{fx}= 
    \begin{pmatrix}
    \cos\left(\frac{\pi x}{2}\right)&\sin\left(\frac{\pi x}{2}\right) \\ -\sin\left(\frac{\pi x}{2}\right)&\cos\left(\frac{\pi x}{2}\right)  
    \end{pmatrix}.
\end{equation}
Then, the generators of the algebra are given by
\begin{align}
    &{\cal Z}_a = (e^{-fX})\indices{_a^b}\Pi_b = 
    \begin{pmatrix}
    \cos\left(\frac{\pi X}{2}\right) \Pi_y - \sin\left(\frac{\pi X}{2}\right) \Pi_z \\
    \sin\left(\frac{\pi X}{2}\right) \Pi_y + \cos\left(\frac{\pi X}{2}\right) \Pi_z
    \end{pmatrix}, \\
    &{\cal X}^a = (e^{f^TX})\indices{^a_b}{\cal X}'^b = \begin{pmatrix}
    \cos\left(\frac{\pi X}{2}\right) Y' - \sin\left(\frac{\pi X}{2}\right) Z'\\
    \sin\left(\frac{\pi X}{2}\right) Y' + \cos\left(\frac{\pi X}{2}\right) Z'
    \end{pmatrix},
\end{align}
and 
\begin{equation}
    {\cal Z}_x = \Pi_x, \qquad {\cal X}^x = X'.
\end{equation}
The only commutator which isn't trivially satisfied is $[{\cal X}^y,{\cal Z}_z]$ (and those related to it). This is 
\begin{align}
    \notag\left[{\cal X}^y(\sigma), {\cal Z}_z(\sigma')\right] =& \left[\cos\left(\frac{\pi X}{2}\right) Y'(\sigma) - \sin\left(\frac{\pi X}{2}\right) Z'(\sigma), \sin\left(\frac{\pi X}{2}\right) \Pi_y(\sigma') + \cos\left(\frac{\pi X}{2}\right) \Pi_z(\sigma')\right] \\
    \notag=& 2\pi i \delta'(\sigma-\sigma')\left(\cos\left(\frac{\pi X(\sigma)}{2}\right)\sin\left(\frac{\pi X(\sigma')}{2}\right) - \sin\left(\frac{\pi X(\sigma)}{2}\right)\cos\left(\frac{\pi X(\sigma')}{2}\right)\right)\\
    \notag=& -2\pi i\delta(\sigma-\sigma')\frac{\pi X'(\sigma)}{2}\left(\cos^2\left(\frac{\pi X(\sigma)}{2}\right)+\sin^2\left(\frac{\pi X(\sigma)}{2}\right)\right)\\
    =& (-2\pi i \delta(\sigma-\sigma')) \left(\frac{\pi}{2}{\cal X}^x(\sigma)\right),
\end{align}
where in the third line we have "integrated by parts". We write the final line in such a way as to make explicit the expected algebra of the elliptic monodromy. The other commutators follow similarly and we thus obtain the full doubled algebra
\begin{align}
    \notag&[{\cal Z}_x(\sigma),{\cal Z}_z(\sigma')] = (-2\pi i \delta(\sigma-\sigma'))\left(\frac{\pi}{2}{\cal Z}_y(\sigma)\right), \qquad [{\cal Z}_x(\sigma),{\cal Z}_y(\sigma')] = (-2\pi i \delta(\sigma-\sigma'))\left(-\frac{\pi}{2}{\cal Z}_z(\sigma)\right), \\
    \notag&[{\cal Z}_x(\sigma),{\cal X}^y(\sigma')] = (-2\pi i \delta(\sigma-\sigma'))\left(-\frac{\pi}{2}{\cal X}^z(\sigma)\right), \qquad [{\cal Z}_x(\sigma),{\cal X}^z(\sigma')] = (-2\pi i \delta(\sigma-\sigma'))\left(\frac{\pi}{2}{\cal X}^y(\sigma)\right), \\
    &[{\cal X}^y(\sigma),{\cal Z}_z(\sigma')] = -[{\cal X}^z(\sigma),{\cal Z}_y(\sigma')] = (-2\pi i \delta(\sigma-\sigma'))\left(-\frac{\pi}{2}{\cal X}^x(\sigma)\right),
\end{align}
and the central extension
\begin{equation}
    [{\cal Z}_i(\sigma), {\cal X}^j(\sigma')] = 2\pi i \delta_i^j \delta'(\sigma-\sigma'),
\end{equation}
agreeing with the doubled geometry \cite{Hull:2009sg}. This background is interesting because it is a genuine string theory background that can be obtained from the supergravity construction as a minimum of the potential \cite{Dabholkar:2002sy}. Normally, such minima only satisfy the supergravity equations of motion, but in this case the minimum is equivalent to a toroidal orbifold and is therefore a solution of the full string theory equations of motion. 
\section{Details of commutation relations in section \ref{s: beyond torus bundles}} \label{a: details of commutation relations}

We present here the details of the calculations of the commutation relations of the doubled algebra and the associators in \ref{ss: flux compactifications on a twisted torus}. We considered there the most general geometric flux compactifications.

\subsection{Commutation Relations}
We compute the commutation relations of the generators \eqref{twistedTorusGenerators}. Firstly, for the "$[\cal Z, \cal Z]$" commutators, we have
\begin{align}
    \notag [{\cal Z}_a(\sigma), {\cal Z}_b(\sigma')] =& [e\indices{_a^i }(\sigma)(\Pi_i (\sigma) - B_{i j }(\sigma)X'^j (\sigma)), e\indices{_b^k }(\sigma')(\Pi_k (\sigma') - B_{k l }(\sigma')X'^l (\sigma'))] \\
    \notag =& e\indices{_b^k }(\sigma')\partial_k  e\indices{_a^i }(\sigma)e\indices{_i ^e}(\sigma){\cal Z}_e(\sigma)2\pi i\delta(\sigma-\sigma') \\
    \notag &- e\indices{_a^i }(\sigma)\partial_i  e\indices{_b^k }(\sigma')e\indices{_k ^e}(\sigma'){\cal Z}_e(\sigma')2\pi i\delta(\sigma-\sigma') \\
    \notag &+ 2\pi i\delta(\sigma-\sigma')e\indices{_a^i }(\sigma)e\indices{_b^k }(\sigma')X'^j (\sigma')(-\partial_i  B_{j k }-\partial_k  B_{i j }-\partial_j  B_{k i }) \\
    \notag =& 2e\indices{_a^i }e\indices{_b^j }\partial_{[i }e\indices{_{j ]}^c}{\cal Z}_c(\sigma)2\pi i \delta(\sigma-\sigma')\\
    \notag & -2\pi i\delta(\sigma-\sigma')(K_{abc}{\cal X}^c(\sigma')) \\
    =& -f_{ab}^c{\cal Z}_c(\sigma)2\pi i\delta(\sigma-\sigma') -K_{abc}{\cal X}^c(\sigma')2\pi i\delta(\sigma-\sigma'),
\end{align}
where, for example, we've used $[\Pi_k(\sigma'),e\indices{_a^i}(\sigma)] = -2\pi i\delta(\sigma-\sigma') \partial_k e\indices{_a^i}(\sigma)$.\\~\\
Next, for the "$[\cal Z, \cal X]$" commutators, we have
\begin{align}
    \notag [{\cal Z}_a(\sigma), {\cal X}^b(\sigma')] =& [e\indices{_a^i }(\sigma)\left(\Pi_i (\sigma)-B_{i j }(\sigma)X'^j (\sigma)\right), e\indices{^b_k }(\sigma')X'^k (\sigma')] \\
    \notag =&-e\indices{_a^i }(\sigma)\partial_i  e\indices{^b_k}(\sigma') X'^k (\sigma')2\pi i \delta(\sigma-\sigma')+e\indices{_a^i }(\sigma)e\indices{^b_i} (\sigma')2\pi i\delta'(\sigma-\sigma') \\
    \notag =& -e\indices{_a^i }(\sigma)\partial_i  e\indices{^b_k}(\sigma') X'^k (\sigma')2\pi i \delta(\sigma-\sigma')\\
    \notag&+e\indices{_a^i}(\sigma)\left(e\indices{^b_i}(\sigma)+(\sigma'-\sigma){e\indices{^b_i}}'(\sigma)+O((\sigma-\sigma')^2)\right)2\pi i\delta'(\sigma-\sigma') \\
    =& f_{ac}^b {\cal X}^c(\sigma)2\pi i\delta(\sigma-\sigma')+\delta_a^b2\pi i\delta'(\sigma-\sigma')+O((\sigma-\sigma')^2),
\end{align}
where the higher order terms are proportional to (for $n\geq 2$)
\begin{align}
    \notag \delta'(\sigma-\sigma')(\sigma-\sigma')^n =& \delta'(\sigma-\sigma')\sum_{m=0}^n (\sigma)^{n-m}(-\sigma')^m\begin{pmatrix}
        n \\ m
    \end{pmatrix}  \\
    \notag =& \sum_{m=0}^n \begin{pmatrix}
        n \\ m
    \end{pmatrix}
    (\sigma)^{n-m} (-1)^m\left[(\sigma)^m\delta'(\sigma-\sigma')+m(\sigma)^{m-1}\delta(\sigma-\sigma')\right] \\
    =& (\sigma-\sigma)^n\delta'(\sigma-\sigma')+ n(\sigma-\sigma)^{n-1}\delta(\sigma-\sigma') = 0,
\end{align}
and so
\begin{align}
    [{\cal Z}_a(\sigma), {\cal X}^b(\sigma')] &=  f_{ac}^b{\cal X}^c(\sigma)2\pi i\delta(\sigma-\sigma')+\delta\indices{_a^b}2\pi i\delta'(\sigma-\sigma').
\end{align}
The final "$[{\cal X}, {\cal X}]=0$" commutator follows trivially. We thus obtain the algebra \eqref{twistedTorusAlgebra}, as claimed.

\subsection{Associativity}

Here we verify the nested commutators \eqref{twistedTorusAssociatorZZZ} and \eqref{twistedTorusAssociatorZZX} required to compute the associators. For "$[\cal Z, [\cal Z, \cal Z]]$", we have
\begin{align}
    \notag[{\cal Z}_a(\sigma),[{\cal Z}_b(\sigma'), {\cal Z}_c(\sigma'')]] =& [{\cal Z}_a(\sigma), -f\indices{_{bc}^d}{\cal Z}_d(\sigma'')-K_{bcd}{\cal X}^d(\sigma'')]2\pi i \delta(\sigma'-\sigma'')\\
    \notag =& -4\pi^2 \delta(\sigma-\sigma'')\delta(\sigma'-\sigma'')\left\{f\indices{_{bc}^d}f\indices{_{ad}^e}{\cal Z}_e(\sigma'')+f\indices{_{bc}^d}K_{ade}{\cal X}^e(\sigma'')\right. \\
    &\left. - f\indices{_{ae}^d}K_{bcd}{\cal X}^e(\sigma'')\right\} + 4\pi^2 \delta'(\sigma-\sigma'')\delta(\sigma'-\sigma'')K_{bcd}\delta_a^d,
\end{align}
as claimed. Note that, after \eqref{A1}, it is claimed that the central extension terms vanish when we add the cyclic permutations via delta function manipulations. Specifically, we use
\begin{align}
    \notag\delta'(\sigma-\sigma'')\delta(\sigma'-\sigma'') =&\partial_\sigma\left(\delta(\sigma-\sigma'')\delta(\sigma'-\sigma'')\right)\\
    \notag =&\partial_\sigma\left(\delta(\sigma'-\sigma)\delta(\sigma''-\sigma)\right)\\
    =& \delta'(\sigma-\sigma')\delta(\sigma''-\sigma)+\delta(\sigma-\sigma')\delta'(\sigma-\sigma'').
\end{align}
Substituting this into \eqref{A1}, we see that the central extension term does indeed vanish.\\~\\
For the "$[\cal Z, [\cal Z, \cal X]]$" nested commutator, we have
\begin{align}
    \notag [{\cal Z}_b(\sigma'),[{\cal Z}_c(\sigma''), {\cal X}^a(\sigma)]] =& [{\cal Z}_b(\sigma'), f\indices{_{cd}^a}{\cal X}^d(\sigma)2\pi i \delta(\sigma''-\sigma)+\delta_c^a2\pi i \delta'(\sigma''-\sigma)] \\\notag
    =&-4\pi^2\delta(\sigma'-\sigma)\delta(\sigma''-\sigma)\left(f\indices{_{cd}^a}f\indices{_{be}^d}{\cal X}^e(\sigma)\right) \\&+ 4\pi^2 \delta'(\sigma-\sigma')\delta(\sigma''-\sigma)f\indices{_{cb}^a}.
\end{align}
Other calculations required to compute the associators proceed along similar lines.

\end{document}